\documentclass[twocolumn]{aa}
\usepackage{graphicx}
\usepackage{amssymb}
\usepackage{amsmath}
\usepackage{float}
\usepackage{txfonts}
\usepackage{gensymb}
\usepackage{hyperref}
\usepackage{subfig}
\usepackage{threeparttable}
\usepackage[normalem]{ulem}
\usepackage[figuresright]{rotating}
\usepackage{longtable}
\usepackage{titlesec}

\begin{document}
	
\authorrunning{Qingshun Hu et al}

\titlerunning{The morphological coherence of open clusters}
	
\title{Exploration of morphological coherence in open clusters with a ``core-shell'' structure}

\author{Qingshun Hu \inst{1} \and Yu Zhang \inst{2,4} \and Songmei Qin \inst{3,4} \and Jing Zhong \inst{3} \and Li Chen \inst{3,4} \and Yangping Luo \inst{1}}

\institute{School of Physics and Astronomy, China West Normal University, No. 1 Shida Road, Nanchong 637002, People's Republic of China, (\email{qingshun0801@163.com} \label{inst1}) \and Xinjiang Astronomical Observatory, Chinese Academy of Sciences, No. 150, Science 1 Street, Urumqi, Xinjiang 830011, People's Republic of China, \and Key Laboratory for Research in Galaxies and Cosmology, Shanghai Astronomical Observatory, Chinese Academy of Sciences, 80 Nandan Road, Shanghai 200030, People's Republic of China, \and School of Astronomy and Space Science, University of Chinese Academy of Sciences, No. 19A, Yuquan Road, Beijing 100049, People's Republic of China
	}
	
\date{Received 1 August 2023 / Accepted 30 April 2024}

\abstract{The morphology of open clusters plays a major role in the study of their dynamic evolution. The study of their morphological coherence, namely, the three-dimensional (3D) difference between the inner and outer morphologies of open clusters, allows us to obtain a better understanding of the morphological evolution of open clusters.}
{We aim to investigate the morphological coherence of 132 open clusters with up to 1~kpc from the Sun in the three-dimensional (3D) space within the heliocentric cartesian coordinate frame. The 132 open clusters have a 3D core-shell structure and conform to the ellipsoidal model, with all of them coming from a catalog of publicly available clusters in the literature.}
{We employed the ellipsoid fitting method to delineate the 3D spatial structure of the sample clusters, while using the morphological dislocation (MD) defined in our previous work and the ellipticity ratio (ER) of the clusters' inner and outer structures to characterize the morphological coherence of the sample clusters.}
{The results show an inverse correlation between the ER of the sample clusters and the number of their members, indicating that sample clusters with a much more elliptical external morphology than internal shape generally tend to host a large number of members. Meanwhile, a slight shrinking of the MD of the sample clusters with their members' number may shed light on the significant role of the gravitational binding of the sample clusters in maintaining their morphological stability. Moreover, there are no correlations between the MD and ER of the sample clusters and their age. They are also not significantly correlated with the X-axis, the Y-axis, their orbital eccentricities, and the radial and vertical forces on them. However, the ER of the sample clusters displays some fluctuations in the distributions between it and the above covariates, implying that the morphologies of the sample clusters are sensitive to the external environment if sample effects are not taken into account. Finally, the analysis of the 3D spatial shapes of sample clusters with a small ER or a large ER demonstrates that the number of members lays an important foundation for forming a dense internal system for sample clusters. At the same time, the MD of the sample clusters can serve well as an indicator of their morphological stability, which is built upon a certain amount of member stars.
}
{We present a new insight into the morphological coherence of open clusters, attributed to the combination of their gravitational binding capacity and external environmental perturbations.}


\maketitle{}

%%%%%%%%%%%%%%%%% BODY OF PAPER %%%%%%%%%%%%%%%%%%

\section{Introduction}

Open clusters, which are mainly distributed along the Galactic plane, make up an important part of the Milky Way. As a consequence, it is of great significance for the study of the Milky Way, such as tracing the structures of the galactic spiral arms \citep{dias05}. In the evolution of the open clusters, the number of their members must change due to star evolution \citep{port10}, collisional dynamics, and tidal perturbations \citep{krau20}. Their shapes can visually indicate this change. In addition, the morphology of the open clusters may inherit morphological imprints of their formation period \citep{krau20}, which can allow us to follow their formation mechanism. Therefore, the morphology of open clusters plays a vital role in the formation and evolution of open clusters.

The typical shape presented in the clusters is thought to be the core-corona structure \citep[e.g.,][]{arty66, khol69, chen04, zhai17}, referred to as the layered structure in our previous study \citep{hu23}. Based on this structure, in our series of papers \citep{hu21a, hu21b, hu23}, we defined, for the first time, a parameter denoted as the ``morphological dislocation'' of open clusters to assess the stability of the two-dimensional (2D) morphology of the open clusters in the statistical study of their morphological laws and properties. 

As long as open clusters can survive in their embedded phase \citep{lada03}, the physical processes that may change the morphology of the open clusters can be roughly described as follows. As a cluster evolves, the distribution of stellar members in its core is likely modified by star evolution, mass segregation \citep{vesp09, port10, evan22, noor23}, and internal gravitational interaction among its members. Subsequently, stellar evaporation \citep{port10, rein23}, expansion resulted from dynamical evolution \citep{krau20}, possible intrusion of non-native stars \citep{hu22}, and external perturbations \citep{spit58, port10}, for instance, Galactic tidal stripping due to the Galactic potential \citep{hegg03, erns15, mcdi22, ange23}, Galactic differential rotation along with, gravitational perturbations from giant molecular clouds, would alter the spatial structure of its outer region. In the process above, the internal and external morphologies of the open cluster will gradually show differences. In this work we consider whether such differences have an impact on cluster evolution and cluster morphological stability. Numerous efforts have been undertaken to study the morphology of open clusters \citep[e.g.,][]{jean16, arty66, khol69, oort79, pand90, berg01, nila02, chen04, cart04, sant05, khar09, zhai17, dib18, hete19, pang21, tarr21, zhon22}; however, very few have studied the morphological issue mentioned above due to the lack of parameters to assess it. This has limited (to some extent) our comprehensive and in-depth understanding and study of the open clusters.

To address the cluster morphology question, here we define, for the first time, the parameter of the inner and outer differences in cluster morphology. This ``morphological coherence'', namely, the 3D difference between the morphology of the clusters' core region and their external morphology, similar to the morphological dislocation of  \citet{hu21b}) is a new parameter of cluster morphology that can indicate clusters' morphological stability. The morphological coherence of the open clusters can be considered as the morphological stability for individual clusters. In general, the poor morphological stability of the open clusters corresponds to poor morphological coherence, which can be shown visually in the distribution of their member stars. Furthermore, this morphological coherence is constantly changing as the clusters evolve. Therefore, the morphological coherence of the open clusters can lay an important foundation for understanding their evolutionary processes.

In the present work, we aim to explore the morphological coherence of the open clusters with core-shell structure in 3D space up to 1 kpc from the Sun, based on cluster catalogs from the literature. By fitting a 3D ellipsoid to the sample clusters to obtain their fitting parameters and then calculate the morphological coherence, we attempt to investigate the morphological coherence of open clusters in terms of their basic parameters (age and the number of member stars), spatial positions, and external environment. This will further enrich our morphological study and also help to provide observational evidence for tracing the formation and evolution mechanisms of open clusters. In the following, we describe the data and methodology in Sect.~2. We present a summary of our results in Sect.~3. Finally, Sect.~4 summarizes our work and gives some important conclusions.

\section{Data and method}

\begin{figure}
	\centering
	\includegraphics[angle=0,width=86mm]{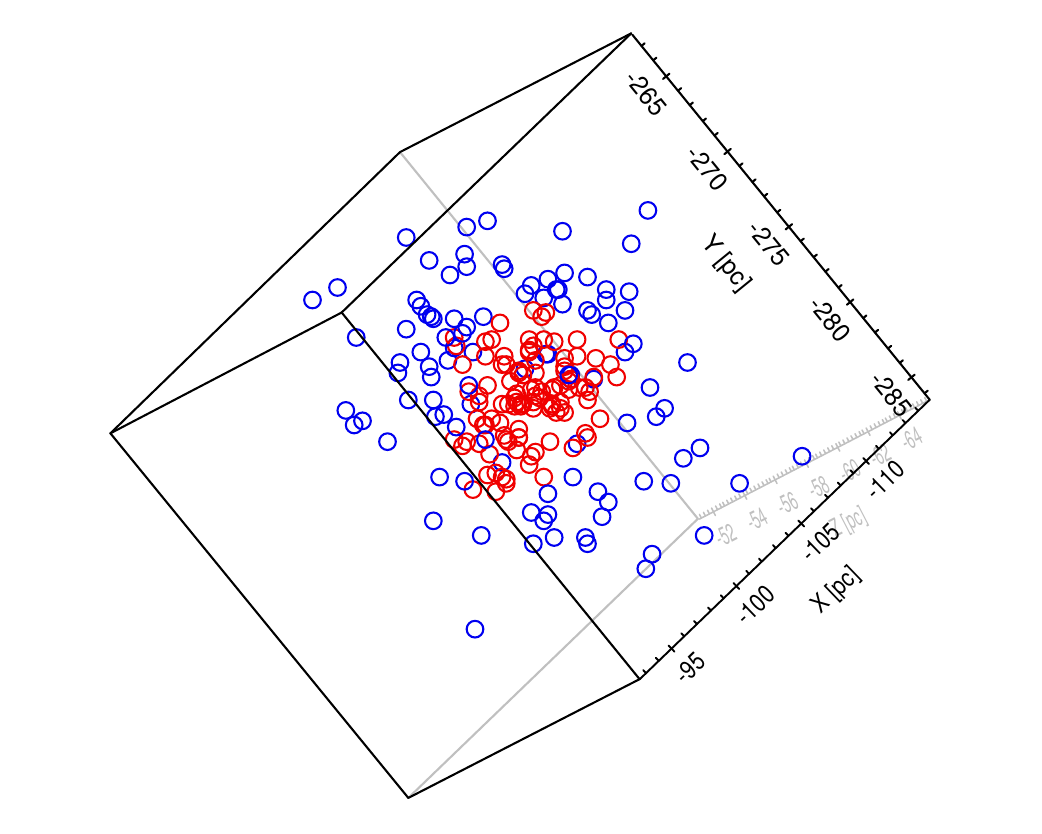}
	\caption{Spatial distribution of the member stars of the cluster (Collinder~135) in the heliocentric Cartesian coordinate system. The red open circles denote the members within the core region of the cluster, with the blue open circles being the stars in its outer region.}
	\label{fig:Collinder_135}
\end{figure}

On the basis of the \textit{Gaia} Data Release~3 (DR3) data, \citet{hunt23} carried out a blind search for member stars of open clusters within all-sky based on the Hierarchical Density-Based Spatial Clustering of Applications with Noise (HDBSCAN) algorithms, with an aim to identify them. They searched 7167 clusters, which included 2387 new candidates and 4105 highly reliable clusters in their catalog, and provided basic parameters and membership lists for all clusters. The sample in this work was selected from their clusters catalog. We filtered 682 sample clusters from the 7167 clusters by six constraints: 1) ``kind''~$=$~``o''; 2) S/Ns~$\geq$~5; 3) class\_50 (median CMD classifications)~$\ge$~0.5; 4) membership probability~$\geq$~0.5; 5) parallax ($\varpi$)~$\geq$~1; 6) members' number (N)~$\geq$~50. The first three terms can ensure the objects we elected are highly reliable open clusters \citep{hunt23}, and the last three conditions can meet the requirements of our sample. It should be noted that the term of member probability~$\geq$~0.5 is able to remove low-quality members \citep{hunt23}, which can guarantee the reliability of member stars. The cut may also remove stars in tidal tails \citep{hunt23}, but not all, so some of our sample clusters still include gravitationally bound and non-gravitational bound member stars. Finally, a 3D ellipsoidal fit was used to filter the sample. The cut criterion is that the errors of the three axes of the fitted ellipsoid for any sample cluster, both core and shell, must be smaller than the length of the three axes, respectively. This guarantees the sample clusters we elected, both core and shell, have stable ellipsoid structures. If the error of any axis of clusters fitted is more than the length of their axis, the fitting ellipsoid structures of the clusters may be unstable or fake. Moreover, the fits were checked visually, frame by frame, to exclude pseudo-ellipsoid fits. These pseudo-ellipsoid fits meet our filter condition, but the member stars of clusters fitted by them is almost exclusively distributed around one patch of the surface of the fitted ellipsoids (especially the ellipsoid fitted of the shell), rather than distributed around the entire surface of the fitted ellipsoids. In this way, we obtained a total of 132 sample clusters, all of which have a ``core-shell'' structure (shown in Fig.~\ref{fig:Collinder_135})  and fit the ellipsoidal model (displayed in Fig.~\ref{fig:Collinder_135_ellipsoid}, with more details given in the next subsection). These 132 open clusters are the research subject of this work.

\subsection{Ellipsoid fitting for the core-shell structure of sample clusters}

The ellipsoid fits were performed on the clusters' core and shell, respectively. The principle formulas for the ellipsoid fitting we adopted can be found in Appendix~\ref{app: ellipsoid fitting method}.

Our research was carried out by fitting the ellipsoid to the core and shell of the sample clusters within the heliocentric Cartesian coordinates frame after correcting the distance of their member stars. This frame refers to XYZ Cartesian coordinates centered on the Sun. The positive X-axis points from the projection of the Sun's position onto the Galactic midplane toward the Galactic center. The positive Y-axis points in the direction of Galactic rotation, with the Z-axis being positive toward the Galactic north pole. The origin of the Cartesian heliocentric coordinate system is the solar system barycenter.

\begin{figure*}
	\centering
	\includegraphics[angle=0,width=140mm]{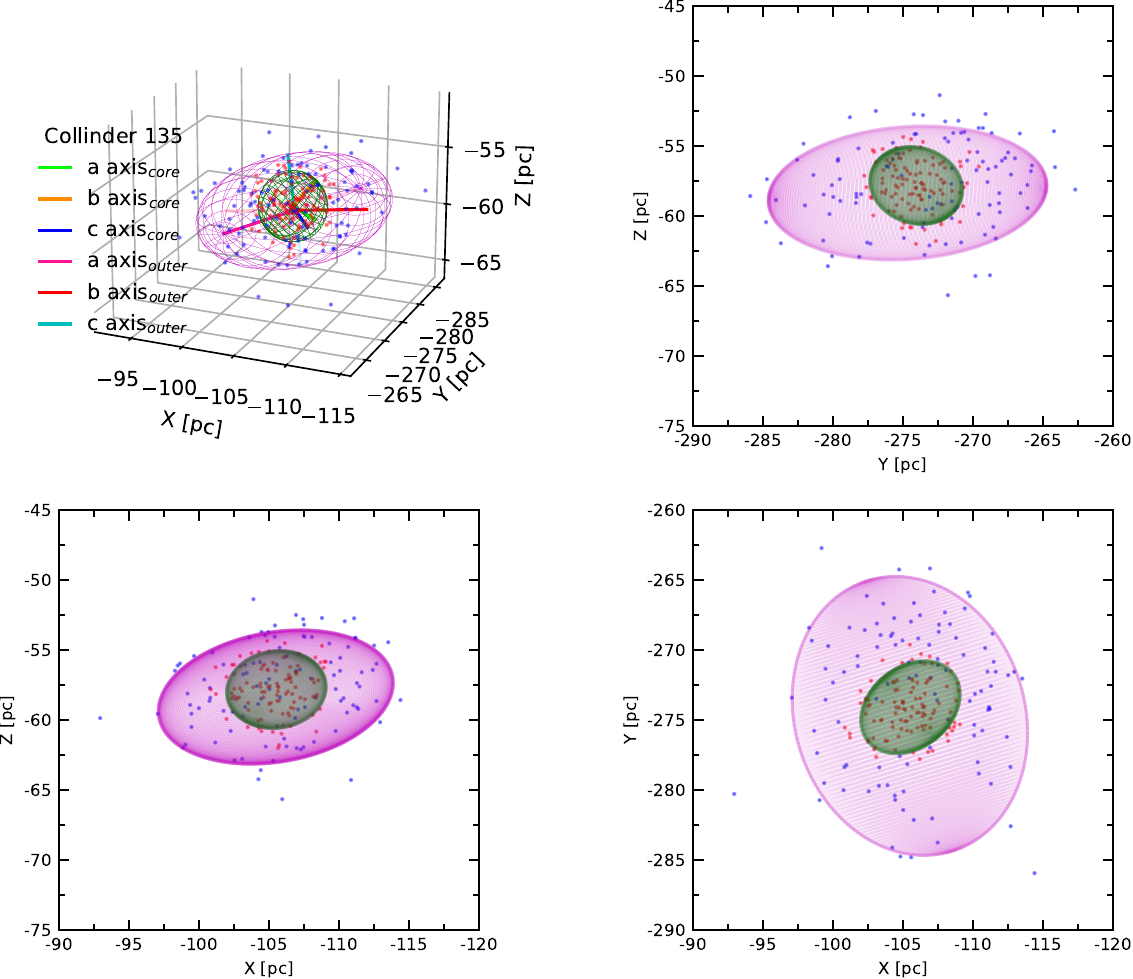}
	
	\caption{3D spatial structure of Collinder~135, with ellipsoidal curves and its three projections. The green ellipsoid with its fitted center position (red pentagram) denotes the core region of the sample cluster, with the purple ellipsoid with its fitted center (blue pentagram) marking its outer structure. The small red and blue dots represent the members of the cluster in the core and outer regions, respectively. The different colored bars represent the different axes of the ellipsoids, respectively, as shown in the legend. The green and purple shaded areas in the three projections (XY plane, XZ plane, and YZ plane) indicate the core projections of the cluster and its shell projections, respectively.}
	\label{fig:Collinder_135_ellipsoid}
\end{figure*}

To fit an ellipsoid to the core-shell structure of sample clusters, we need to first determine the 3D spatial distribution of the clusters' members and then delineate their core-shell structure. Before obtaining the space distribution, we calculated the heliocentric Cartesian coordinates (X, Y, Z) for each star member of the sample clusters. This calculation was performed with the Python Astropy package \citep{astr13, astr18}.

Before calculating the 3D spatial coordinates of the sample's member stars by the Astropy package, we also have to correct the distance of these members. Due to the symmetry of observational errors in \textit{Gaia} measuring parallaxes \citep[see, e.g.,][]{bail15, luri18, carr19, zhan20}, distance errors become an asymmetric distribution function by directly inverting these \textit{Gaia} measuring parallaxes, 1/$\varpi$. Therefore, their 3D spatial distributions appear to be stretched along the line of sight. The distance problem is alleviated within the Bayesian framework \citep[e.g.,][]{bail15, carr19, pang21, ye21, qinm23, hu23}. By this approach, \citet{carr19} has corrected members' distances of NGC~2682, located at about 860 pc from the Sun \citep{cant18}. The principle equations for the Bayesian distance correction are listed in Appendix~\ref{distance correction}.

\begin{figure*}
	\centering
	\includegraphics[angle=0,width=76mm]{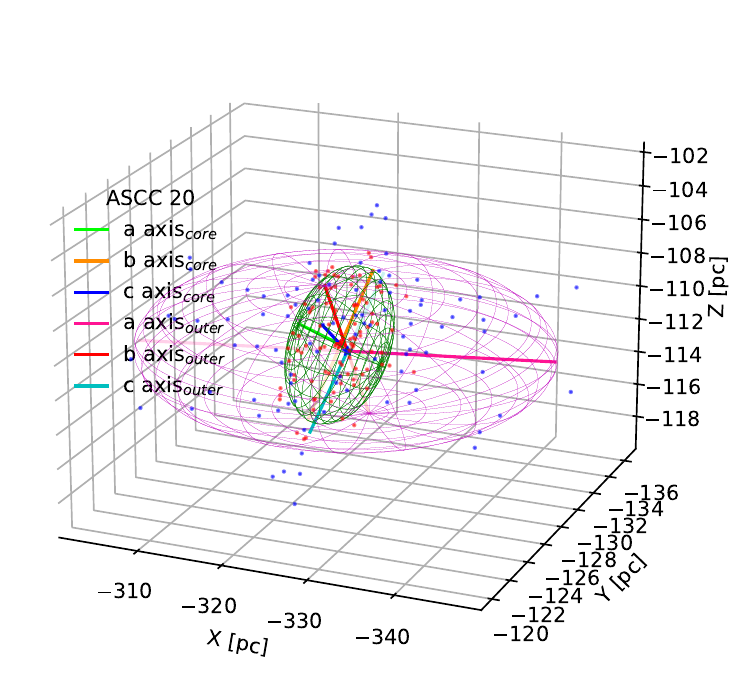}
	\includegraphics[angle=0,width=76mm]{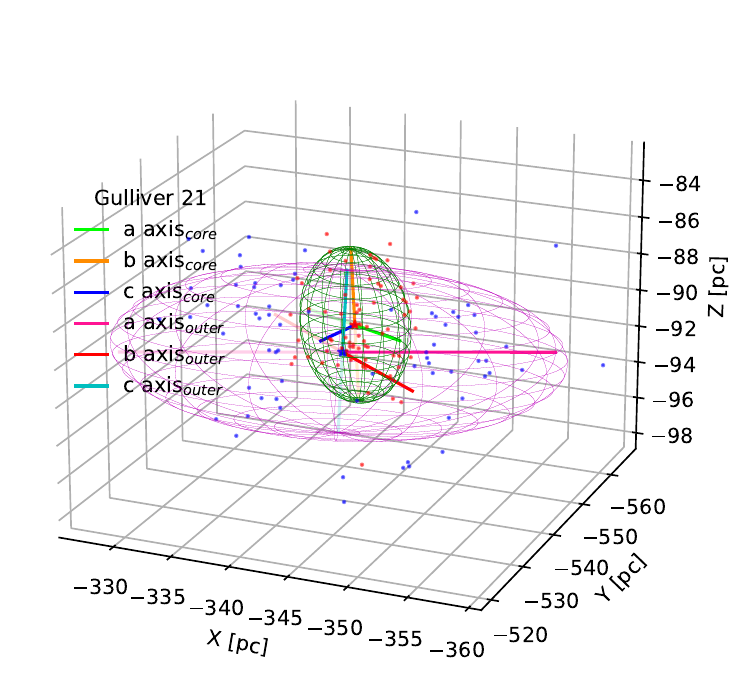}
	\includegraphics[angle=0,width=76mm]{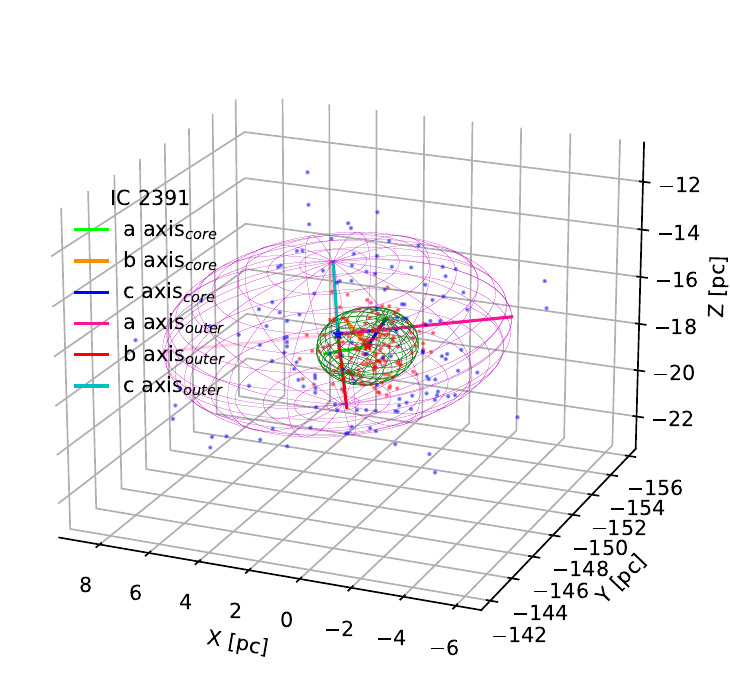}
	\includegraphics[angle=0,width=76mm]{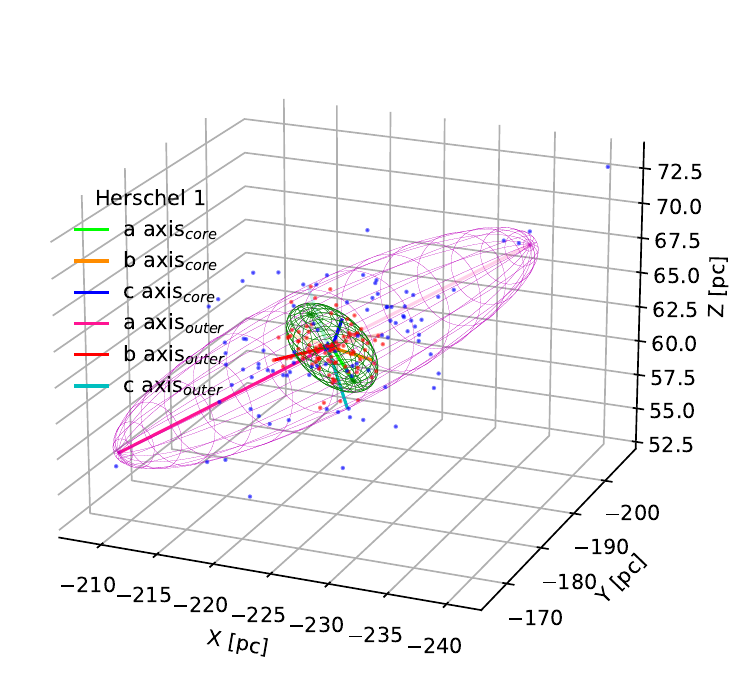}
	\caption{3D spatial structure of the sample clusters (left upper: ASCC~20; right upper: Gulliver~21; left lower: IC~2391; right lower: Herschel~1) and their ellipsoidal curves. The symbols and curves of the picture are the same as those of Fig.~\ref{fig:Collinder_135_ellipsoid}.}
	\label{fig:fourOCs_ellipsoid}
\end{figure*}

In light of the core-shell structure of star clusters and the layered structure of open clusters detected by \citet{hu23}, we divided the stellar members of each sample cluster into two different regions (including a core and outer regions, i.e., the core represents those stars within the radius of the half number, with the outer field indicating those members out of this radius) based on the radius of the half number of the cluster member stars (shown in Fig.~\ref{fig:Collinder_135}). The starting point of this half-number radius is the centroid of the 3D spatial distribution of the member stars, which is obtained from the 3D nuclear density assessment after the members' distance correction. We set out to delineate the morphologies of the two different regions of each sample cluster in 3D space by ellipsoid fitting.  The fitting ellipsoids of the sample clusters thus were obtained, as shown in Fig.~\ref{fig:Collinder_135_ellipsoid} and \ref{fig:fourOCs_ellipsoid}.

\subsection{Characterization of the morphological coherence of the sample clusters}

The morphological coherence of the sample clusters, in fact, refers to a 3D difference between the inner and outer morphologies in this work. It is essentially determined by their core-shell structures, mainly in the 3D spatial distributions of member stars in their inner and outer regions. Any variable that could measure this disparity between the inner and outer regions can indicate the cluster's morphological coherence. The morphological dislocation is probably an appropriate indicator. The morphological dislocation of open clusters (hereafter MD), namely, the offset between the centers of the core and outer regions, is important in the study of their layered structure and morphological stability. Its definition was originally taken from our previous paper of this series \citep{hu21b}, which was performed in 2D space. In this work, we employ it to probe the morphological coherence of the clusters. Unlike our previous work, here it was operated in three dimensions. To clearly describe this new definition, the detailed equation is listed below:

\begin{equation}
	d = \frac{\sqrt{(X_{outer} - X_{core})^2 + (Y_{outer} - Y_{core})^2+ (Z_{outer} - Z_{core})^2}}{a_{outer}}.
	\label{equation}
\end{equation}

Here, $d$ is a dimensionless physical quantity denoting the morphological dislocation of the sample clusters in 3D space. The value of $d$ specifically refers to the system centers' relative offset between the core and the outer regions of the clusters. Then, $X_{core}$, $Y_{core}$ and $Z_{core}$ denote the center coordinates of the fitting ellipsoid in the core regions of the clusters (see the solid green curves in Figs.~\ref{fig:Collinder_135_ellipsoid} and \ref{fig:fourOCs_ellipsoid}), while $X_{outer}$, $Y_{outer}$ and $Z_{outer}$ refer to the center coordinates of the fitting ellipsoid in their outer regions (see the solid purple curves in Figs.~\ref{fig:Collinder_135_ellipsoid} and \ref{fig:fourOCs_ellipsoid}). $a_{outer}$ represents the semi-major axis of the fitted ellipsoid of their outer fields. The error of $d$ is calculated based on error propagation. The left panel of Fig.~\ref{fig:error} shows the distribution between $d$ (MD) and its error ($\sigma_{d}$).

To fully explore the morphological coherence within the sample clusters, we also took the ellipticity ratio ($e_{core}$/$e_{outer}$) of their core-shell structures (hereafter, ER) to describe the morphological difference between their inner and outer structures, characterizing the morphological coherence of the clusters. The detailed ellipticity formula is as follows:

\begin{equation}
	\qquad \qquad  \qquad  \qquad e_{core} = 1 - \frac{c_{core}}{a_{core}}
,\end{equation}

\begin{equation}
	\qquad \qquad  \qquad  \qquad e_{outer} = 1 - \frac{c_{outer}}{a_{outer}}
,\end{equation}

Here, $e_{core}$  and $e_{outer}$ refer to the ellipticities of the core and shell regions of the sample clusters, respectively; $a_{core}$ and $c_{core}$ represent the semi-major and semi-minor axes of the fitted ellipsoids of their core fields, respectively, while $a_{outer}$ and $c_{outer}$ correspond to the semi-major and semi-minor axes of the fitted ellipsoids of their outer regions, respectively. The errors of $e_{core}$, $e_{outer}$, and $e_{core}$/$e_{outer}$ are also calculated according to error transmission. The right panel of Fig.~\ref{fig:error} displays the distribution between $e_{core}$/$e_{outer}$ (ER) and its error ($\sigma_{e_{core}/e_{outer}}$).

When the ER ($e_{core}$/$e_{outer}$) is greater than 1 or less than 1, the inner and outer morphologies of the sample clusters are different. It is noted that the ER greater than 1 indicates that the core of the clusters is more elliptical than the shell and, conversely, the ER less than 1 suggests that the core of the sample clusters is less elliptical compared to their shell. \citet{ange23} pointed out that clusters tend to form a compact core along their dynamical evolution process, which may be a result of the movement of massive members toward the clusters' core based on two-body interactions. Meanwhile, the sphericity of the cores also increases \citep{chen04}. The sample clusters with ER less than 1 or greater than 1 might face two scenarios. Clusters with more elliptical external morphology than internal morphology likely have undergone a two-body relaxation process \citep{port10}, tending to have a circular internal structure \citep{hegg03, chen04} and an advanced evolution status \citep{ange23}. In addition, this shape presents likely a large Roche volume filling factor \citep{ange23} representing the degree of tidal filling of a cluster \citep{sant20, ange23}, which means that the clusters with this morphology are subject to both external tidal fields and internal evolution. On the contrary, the sample clusters with an elliptical core and a circular shell may be suffering from an internal two-body relaxation.

%%%%%%%%%%%%%%%%%%%%%%%%%%%%%%%%%%%%%%%%%%%%

\section{Result}

\subsection{Distribution of morphological coherence of sample clusters}

We obtained all parameters, including the MD and ER of 132 sample clusters, through ellipsoid fitting and expressions above defined, compiled a machine-readable format and seen in Appendix~\ref{table:parameters}. Figure~\ref{fig:error} displays the distributions of the MD with its error (left panel) as well as the ER with its error (right panel). We found that the errors of the MD of most sample clusters are less than 0.1, with most ER errors less than 0.2. Meanwhile, both errors, expressed as $\sigma_{d}$ and $\sigma_{e_{core}/e_{outer}}$, respectively, present relatively apparent positive correlations with the MD and ER. The main reason is that the fitting error is larger for the sample clusters with large MD or large ER.

\begin{figure*}
	\centering
	\includegraphics[angle=0,width=72mm]{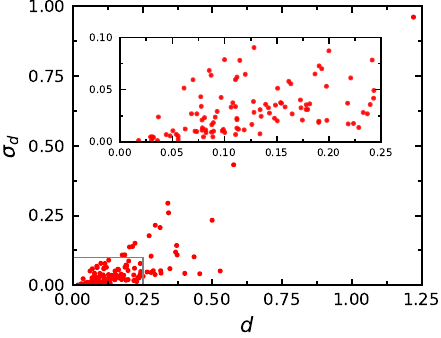}
	\includegraphics[angle=0,width=72mm]{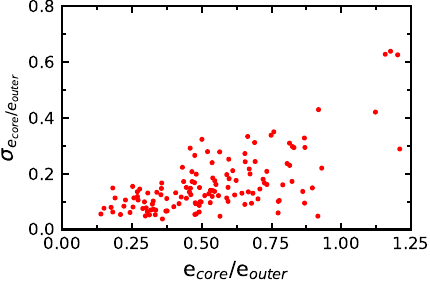}
	\caption{Scatter diagrams of the morphological dislocations (MD) and the ellipticity ratios (ER) of the sample clusters with their errors. Each red dot in the left and right panels denotes one sample cluster, with a zoom region (gray rectangular) in the left subplot.}
	\label{fig:error}
\end{figure*}

\begin{figure}
	\centering
	\includegraphics[angle=0,width=86mm]{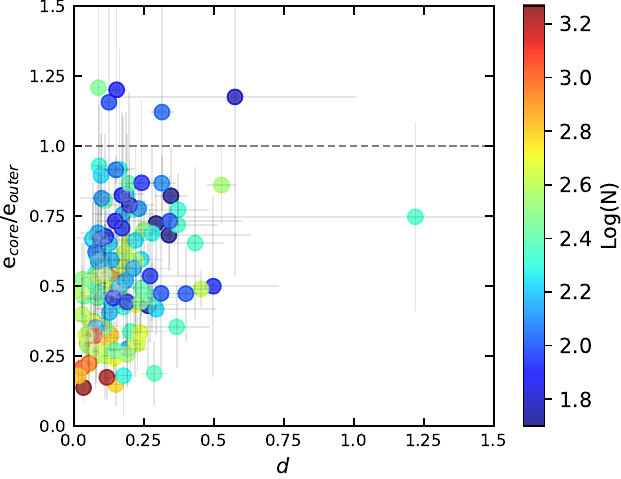}
	\caption{Distribution of the MD of the sample clusters and their ER. Each colored circle represents one sample cluster, with its color being coded according to the logarithmic value of the number of member stars of the sample clusters. The horizontal and vertical gray bars indicate the errors in the MD of the clusters and their ER, respectively. The dashed line denotes the position of the smallest morphological difference between the clusters' inner and outer regions.}
	\label{fig:ratio~d}
\end{figure}

To get a preliminary view of the morphological coherence distribution of the sample clusters, we plotted the distribution between the MD and the ER of the sample clusters, as shown in Fig.~\ref{fig:ratio~d}. Each colored circle represents one sample cluster, with its color representative of the number of the sample clusters' members. The gray bars (horizontal and vertical directions) cross it are the error bars of MD and ER, respectively. The horizontal dashed line in the Figure marks the position of ER equal to 1, where the ellipticity of the cluster's internal morphology is consistent with the ellipticity of its external morphology. The picture shows an increasing tendency of the ER ($e_{core}$/$e_{outer}$) values as a function of the MD ($d$), which implies that sample clusters with more elliptical external morphology than internal morphology typically present smaller morphological dislocations (MD), as shown by the dark red-filled circles in Fig.~\ref{fig:ratio~d}; thus, they have greater morphological stability. Meanwhile, it is apparent that these clusters have a larger number of member stars. Therefore, we concluded that the number of sample clusters' members plays a crucial role in keeping their morphological stability.

\begin{figure*}
	\centering
	\includegraphics[angle=0,width=80mm]{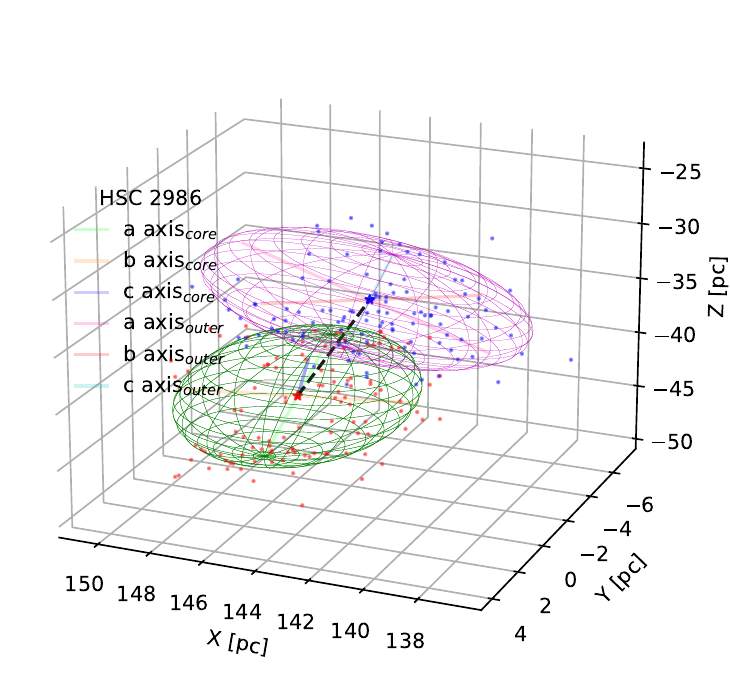}
	\includegraphics[angle=0,width=80mm]{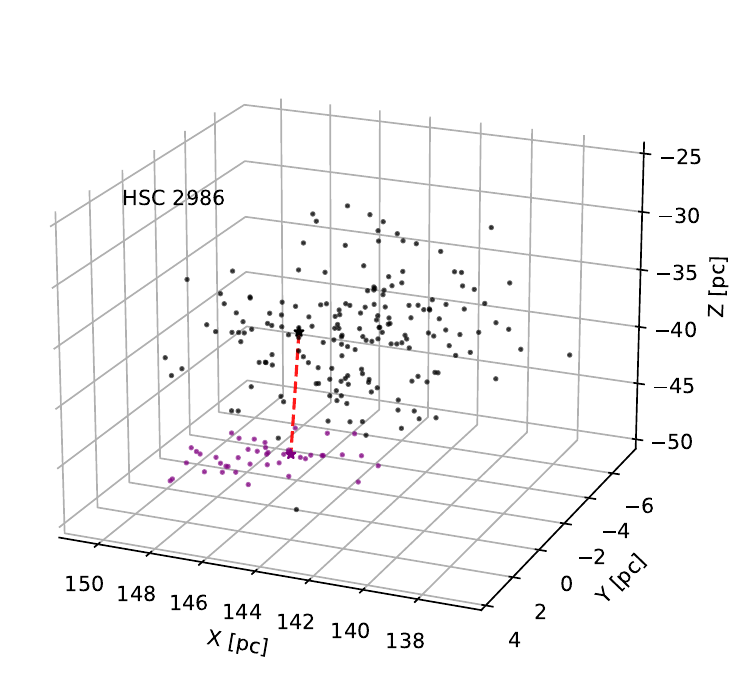}
	\caption{3D spatial structure of HSC~2986 and its ellipsoidal curves (left). The black dashed line indicates the unnormalized MD of HSC~2986. Other symbols and curves are the same as those in Fig.~\ref{fig:Collinder_135_ellipsoid}. 3D spatial structure of HSC~2986 (right). The purple and black dots are divided by the Gaussian mixture model (GMM) method, with the red dashed line meaning the unnormalized MD of HSC~2986. The purple and black pentagrams are the maximum points of the density of the purple and black scatter distribution.}
	\label{fig:HSC_2986}
\end{figure*}

In addition, we found (as seen in Fig.~\ref{fig:ratio~d}) a sample cluster (HSC~2986) that departs from the distribution of most sample clusters. It is a special object for our sample population because it has a relatively small number of members and the largest MD among all samples, although the error of the cluster's MD is large. To examine the cluster further, we traced its 3D spatial distribution, as shown in Fig.~\ref{fig:HSC_2986}. The left panel of the Figure presents the 3D spatial distributions and the fitted ellipsoids of the internal and external structures of the HSC~2986. Its internal and external structures were divided based on the radius of the half number of cluster members. In this panel, the black dashed line indicates the position of the MD of HSC~2986, with its length representing, the unnormalized MD (see Eq.~\ref{equation}). To verify whether the cluster really has a large unnormalized MD, we adopted the Gaussian mixture model (GMM) used by \citet{tarr21} to distinguish among open clusters' core region, tidal tail, and corona in two dimensions, with 1000 iterations to divide the members' distribution of HSC~2986 into two components, as shown in the right panel of Fig.~\ref{fig:HSC_2986}. The two components are purple and black dots in this picture. We then evaluated the density centers (purple pentagram and black pentagram) in the two parts of the cluster separately, finally, we derived its unnormalized MD (see the red dashed line in the right panel of Fig.~\ref{fig:HSC_2986}).

By comparing the unnormalized MDs of HSC~2986 obtained, we are able to determine that HSC~2986 has a large unnormalized MD, which implies that the cluster is weakly morphologically stable. Moreover, we can see that the whole members' distribution of HSC~2986 does not show a distinct core, presenting instead a diffuse spatial arrangement. Therefore, we speculated that the cluster may be susceptible to disruption by external forces, regardless of the cluster's evolutionary state. However, it is also possible that its core and outer structures are two separate groups due to the large unnormalized MD if HSC~2986 is misclassified as an open cluster.

\subsection{Correlations of the morphological coherence of sample clusters with the number of members and age}

\label{Sec:coherence}

The morphological coherence (MD and ER) of sample clusters can be regarded as an indicator of their morphological stability. Theoretically, the morphological stability is in a struggle between the clusters' internal self-constraints and external environmental perturbations. In this section, we intend to explore the internal influence factors on the morphological coherence of the sample clusters.

\begin{figure*}
	\centering
	\subfloat[Distributions of the MD and ER of sample clusters with their members' number.]{
		\centering
		\includegraphics[width=76mm]{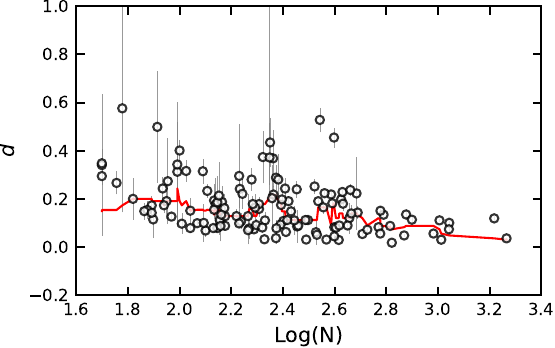}
		\includegraphics[width=76mm]{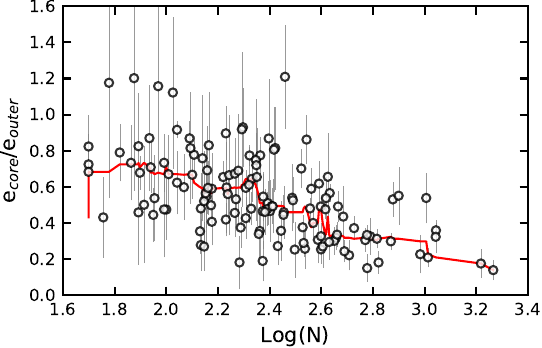}
	}\\
	\subfloat[Distributions of the MD and ER of sample clusters with their ages.]{
		\centering
		\includegraphics[width=76mm]{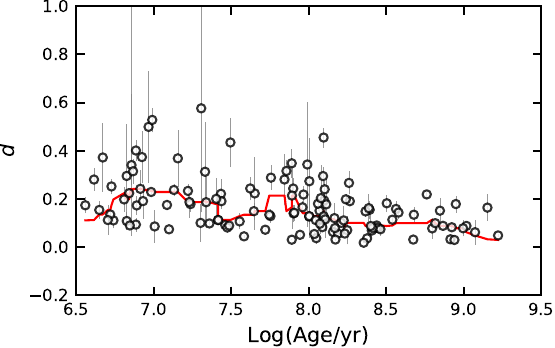}
		\includegraphics[width=76mm]{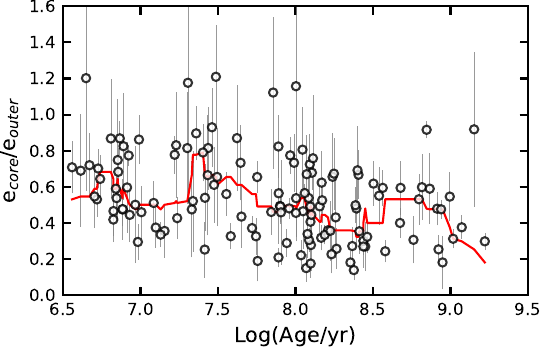}	
	}
	\caption{Morphological coherence (MD and ER) versus the fundamental parameters. The open circles in all panels represent the sample clusters, with the black bars being the errors of the morphological coherence (MD and ER). The red curves indicate the median filter lines of all distributions.}
	\label{fig:correlation_Nage}
\end{figure*}

Figure~\ref{fig:correlation_Nage} displays the distributions between the morphological coherence (MD and ER) and the number of the clusters' members as well as their ages, with their median filter curves. It is apparent (as seen in Fig.~\ref{fig:correlation_Nage}a) that as the number of clusters' members increases, both their MD and ER show decreasing trends; among them, the anti-correlation between the ER of sample clusters and the number of their members is more significant. This indicates that the members' number of the sample clusters has a significant impact on their ER and the sample clusters with a large number of members have much less elliptical internal morphology than external morphology. The result of a shrinking of the MD of sample clusters with the number of their members can be interpreted as the influence of self-gravitational binding of sample clusters on their morphologies, because the larger the number of member stars in a sample cluster, the stronger its self-gravitational binding capacity.

Moreover, there may be a compact internal structure in a cluster with a large number of members. This structure is prone to smaller dynamical timescales, thereby speeding up the cluster evolution, including the evaporation rates and the process of mass segregation \citep{spit69, madr12}. It is well known that member stars in the evaporation process escape almost from the external region within the tidal radius of clusters through Lagrangian points of the Jacobi potential sphere \citep[e.g.,][see respectively their Fig.~6 and Fig.~15]{port10, krum19}. The Jacobi potential sphere of the cluster with a number of members that looks like an egg may be fully filled by its members, while its internal morphology tends to be compact and spherical due to two-body interaction \citep{chen04}. The analysis above suggests that the number of the member stars of sample clusters is most likely an important factor in maintaining their morphological coherence, implying that the gravitational binding of the clusters is responsible for keeping their structure stable and less susceptible to external perturbations.

In addition, we explored whether age affects the morphological coherence of the sample clusters. As open clusters age, their member stars redistribute as a result of the two-body relaxation process, leading to morphological changes in their internal structures. It has been theoretically speculated that the MD of the sample clusters varies with their ages. Older clusters are still alive. Apart from the disturbance of the external environment, the morphological stability of the clusters is most likely a talisman for their stable survival. In this way, we explored the relationship between the MD and ER of the sample clusters and their ages, as shown in the (b) subplot of Fig.~\ref{fig:correlation_Nage}.

From the left panel of the (b) subplot of Fig.~\ref{fig:correlation_Nage}, we can see that as the age of sample clusters increases, their MD presents only a slight downward trend, or even no change if we are looking at fluctuations of the median filter line (red). This demonstrates that the MD of the sample clusters is independent of their age, which is inconsistent with our speculation above. This is because the age is only a time scale of clusters existing and evolving. Many clusters may dissolve at any age due to weaker self-gravitational bindings, stronger external perturbations, or collisions with giant molecular clouds. In spite of the external influence on the evolution of the sample clusters, the number of their members is a key factor for their existence. For different clusters with different numbers of members, they may have the same age; thus, we can find their different MDs at the same age.

Not coincidentally, the distribution of ERs of the sample clusters with respect to their ages does not show any significant upward or downward trend, as observed in the right panel of the (b) subplot of Fig.~\ref{fig:correlation_Nage}. However, we can see big local fluctuations (logt~$<8$, see red line) in ER as the age of the sample clusters increases. Although this may be due to sample effects, it is also likely to be consistent with the fact that the young sample clusters are complexly evolving. Therefore, it may appear a change in the internal and external morphologies of these young sample clusters, then resulting in a change in their ERs. Since the sample clusters evolve as they age, they are exposed to multiple-body gravitational interactions with their evolution from both the external environment and internal stellar evolution, so their ERs will be in a fluctuating process all the time.

\subsection{Correlations of the morphological coherence of sample clusters with spatial position}

\label{Sec:coherencetwo}

Theoretically, the morphological stability of the sample clusters must be subject to the disturbance of the external environment. The external environment may vary with different spatial positions of the clusters. The positions we here want to study are within the heliocentric Cartesian coordinates frame. In this section, we intend to explore the relationship between the morphological coherence of the sample clusters and their spatial positions.

\begin{figure*}
	\centering
	\includegraphics[angle=0,width=180mm]{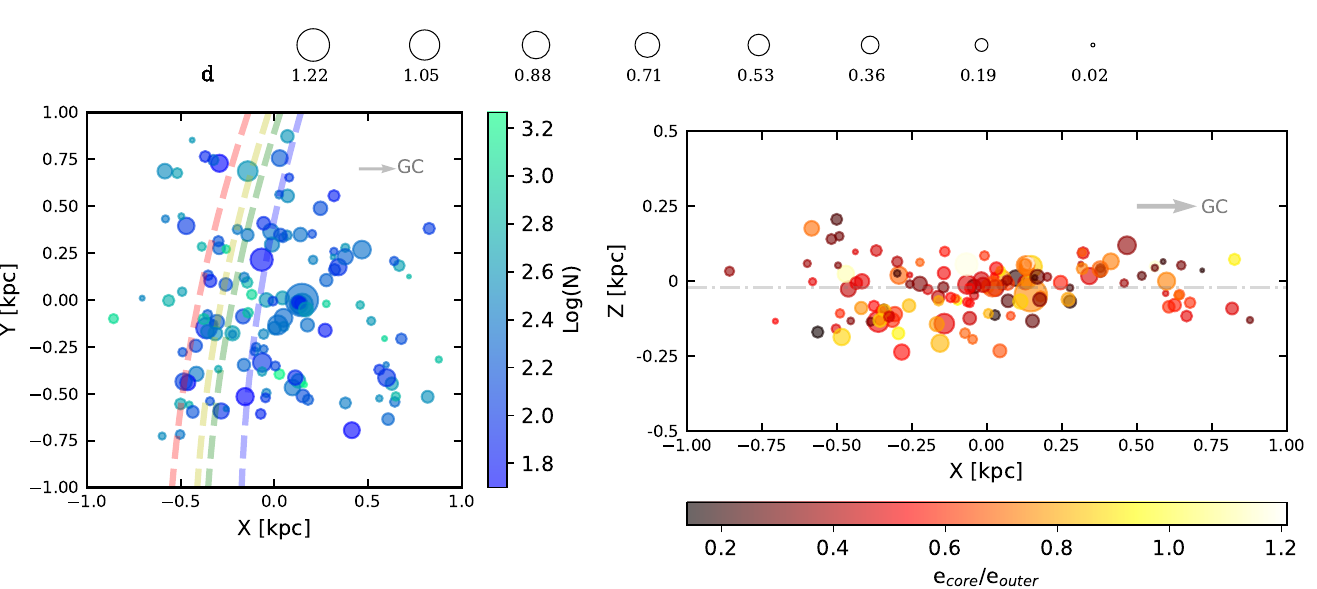}
	\caption{Distribution of the sample clusters on the plane of the galactic disk (left). Each circle represents a sample cluster, with its size proportional to the MD of the sample, and its color is coded according to the members' number of the sample clusters. The different colored dashed lines indicate the positions of the local arm detected by different types of tracers, the data involved was from \citet{hao21}. Distribution of the sample clusters on the profile map of the disk (right). Each circle represents one cluster, with its size being proportional to the cluster's MD and its color coded according to the cluster's ER. The dot-dashed line denotes the mid-plane (Z~$\geq$~-20.8~pc) of the disk.}
	\label{fig:XYZ}
\end{figure*}

Figure~\ref{fig:XYZ} shows the distributions of our sample clusters on the XY plane and XZ plane of the heliocentric Cartesian coordinates frame. At the same time, we also present the MD, the ER, and the number of member stars of the sample clusters in the distributions. Each colored circle in Fig.~\ref{fig:XYZ} represents a sample cluster. Its size is proportional to the MD of the cluster, while its color is coded according to the number of member stars of the cluster (left panel) or its ER (right panel). Four thick dashed lines of different colors in the left panel mark the position of the local arm. These dashed lines are not at the same position because they are defined by different types of tracers \citet{hao21}. The dot-dashed line in the right panel represents the mid-plane of the Galactic disk. We found from the left panel of Fig.~\ref{fig:XYZ} that the local arm spans the distribution of the sample clusters on the XY plane. The sample clusters that are in the domain of the local arm are almost uniformly distributed, as well as the number of their member stars (marked by circles' color), and their MDs (displayed by circles' size), are also almost uniformly distributed on the XY plane. The only caveat is that most of our samples are on the inside of the local arm. In addition, we can see, as shown in the right panel of Fig.~\ref{fig:XYZ} that the distribution of the sample clusters on the XZ plane is within $|Z|$~$\leq$~250~pc and does not show any particular pattern, apart from their aggregation on the mid-plane. Meanwhile, the MD (represented by the circles' size) and ER (coded by the circles' color) of the sample clusters appear to be randomly distributed in the XZ plane.

\begin{figure*}
	\centering
	\subfloat[Distributions of the MD of the sample clusters with their spatial location.]{
		\centering
		\includegraphics[width=76mm]{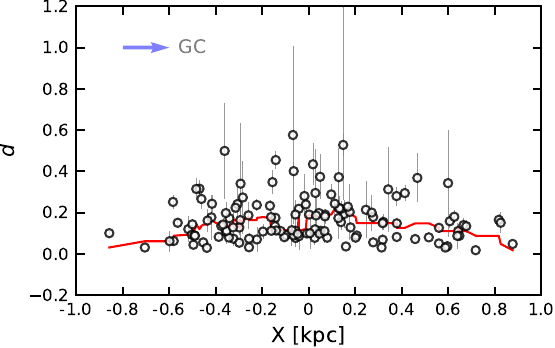}
		\includegraphics[width=76mm]{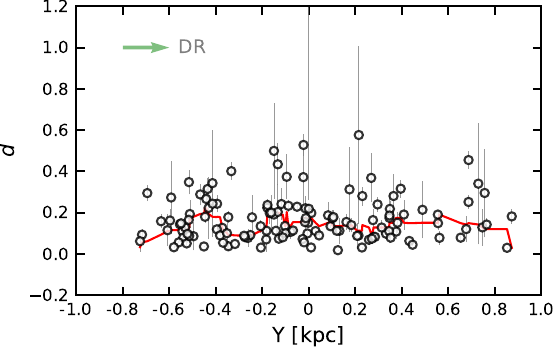}
	}\\
	\subfloat[Distributions of the ER of the sample clusters with their spatial location.]{
		\centering
		\includegraphics[width=76mm]{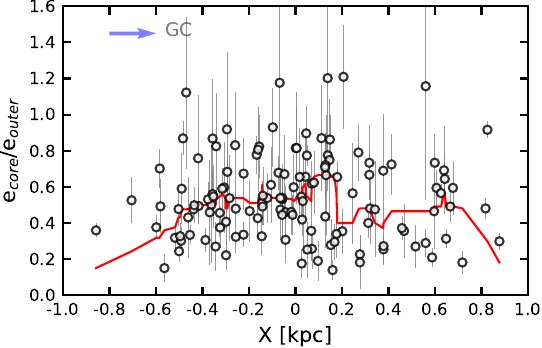}
		\includegraphics[width=76mm]{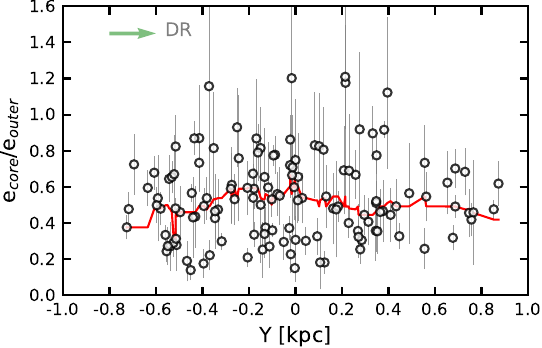}
	}
	\caption{Morphological coherence (MD and ER) versus Spatial position. All open circles (representing the sample clusters), all black error bars (being the errors of the MD or ER), and red curves (denoting median filter lines), same as those in Fig.~\ref{fig:correlation_Nage}. ``GC'' and ``DR'' represent the Galactic center and the Galactic differential rotation, respectively.}
	\label{fig:Coherence_XY}
\end{figure*}

Although the MD and ER of the sample clusters do not show any interesting distribution trends on the XY plane, it is worth exploring their distribution on the X-axis and Y-axis separately, because the morphology of the sample clusters may be affected differently at different spatial locations. Figure~\ref{fig:Coherence_XY} shows the distributions of the MD and ER along the X-axis and Y-axis, respectively. We also plotted their median filter lines (red curves) in this picture. In the left panel of the (a) subplot of Fig.~\ref{fig:Coherence_XY}, we did not find any increasing or decreasing trend between the MD and the X-axis, which implies the MD of the sample clusters being independent of the X-axis. This result is not consistent with our expectations because along the X-axis, the closer to the Galactic center, the more disturbed the morphology of the sample clusters are; hence, the larger their MDs are expected to be. Of course, it is likely due to a small number of sample clusters. Therefore, we need more open clusters to verify this result in the future, if possible. In addition, we also found that the MDs of the sample clusters are evenly distributed along the Y-axis. Since along the Y-axis, the distributions of the sample clusters are almost symmetric in the regions of its positive and negative semi-axes, the MD has nothing to do with the Y-axis.

Furthermore, the ER of the sample clusters presents fluctuations along the X-axis toward the Galactic center, as displayed in the left panel of the (b) subplot of Fig.~\ref{fig:Coherence_XY}, but it does not show any trend at a whole. This means that the ER of the sample clusters is sensitive to the external environment along the X-axis if sample effects are not considered. To say the least, the external morphology of the sample clusters changes along the X-axis due to changes in external gravitational perturbations. The closer we get to the Galactocentric side, the stronger the potential in the external environment and the more perturbation the morphology of the sample clusters will suffer. Like the MDs of the sample clusters, their ER also does not have any significant relationship with the Y-axis, as shown in the right panel of the (b) subplot of Fig.~\ref{fig:Coherence_XY}.

\subsection{Correlations of the MD of sample clusters with external environment}\label{Sec:coherencethree}

\begin{figure*}
	\centering
	\subfloat[Distributions of the eccentricity, F$_{R}$, and F$_{z}$ of the sample clusters with their MD.]{
		\centering
		\includegraphics[width=60mm]{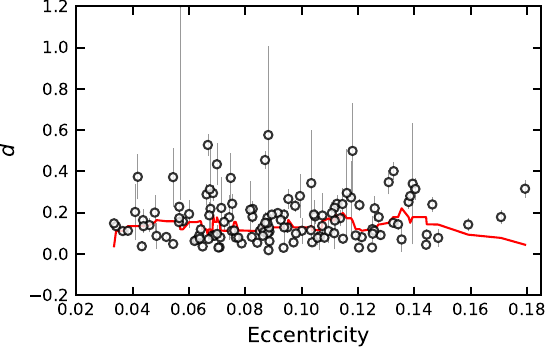}
		\includegraphics[width=60mm]{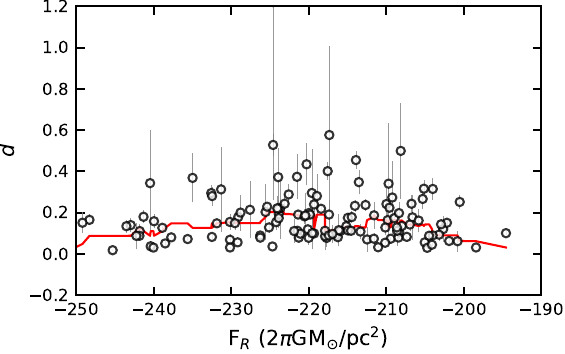}
		\includegraphics[width=60mm]{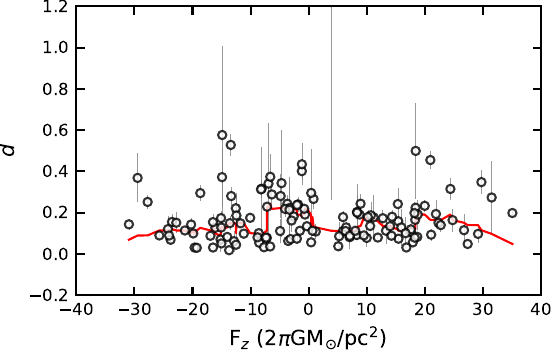}
	}\\
	\subfloat[Distributions of the eccentricity, F$_{R}$, and F$_{z}$ of the sample clusters with their ER.]{
		\centering
		\includegraphics[width=60mm]{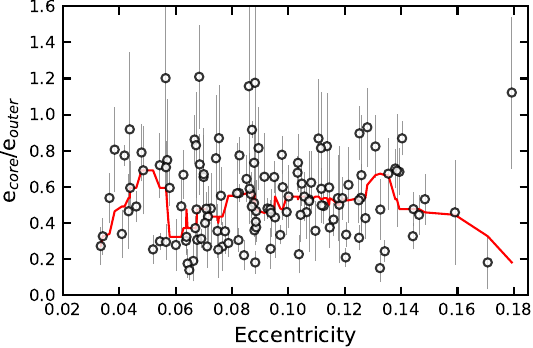}
		\includegraphics[width=60mm]{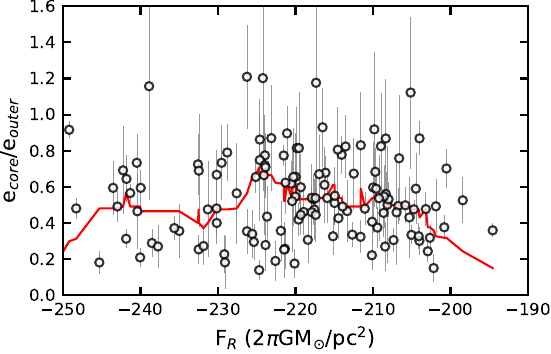}
		\includegraphics[width=60mm]{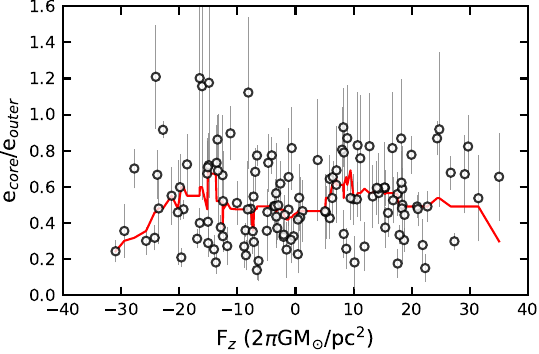}
	}
	
	\caption{Morphological coherence versus external environmental parameters. All open circles (representing sample clusters), all black error bars (being errors of the MD or ER) and red curves (indicating median filter lines) are the same as those in Fig.~\ref{fig:correlation_Nage}.}
	\label{fig:coherencethree}
\end{figure*}

In Section \ref{Sec:coherencetwo}, although the MD of the sample clusters is not related to their spatial positions, the external environment around the positions may be different. Therefore, some forces embedded in the external environment may impact their morphological coherence. To explore it, we employ the \texttt{Python} module \texttt{galpy} \citep{bovy15} to derive some parameters involving the external environment, based on the \texttt{galpy.potential} module: \texttt{MWPotential2014}. The axisymmetric potential module, derived by a simple model fit to the available dynamical data for our galaxy, comprises a Miyamoto-Nagai disk \citep{miya75}, a bulge, and a dark matter halo modeled with a Navarro-Frenk-White (NFW) potential \citep{nava97}. Inputting the coordinate, proper motion, and radial velocity data of the sample clusters, and then we can obtain orbit eccentricities, radial forces (F$_{R}$), and vertical forces (F$_{z}$) in cylindrical coordinates\footnote{It is a cylindrical coordinate centered on Galactic center, following left-handed frame in \texttt{galpy}.} of these sample clusters. The meaning of these parameters can be also found on the web\footnote{\url{https://docs.galpy.org/en/v1.7.0/}}.

Since the sample clusters with large orbital eccentricities necessarily move across different potential energy surfaces in the galactic disk compared to those with small eccentricities, their MD and ER are consequentially affected to a greater extent. In the left panels of the (a) and (b) subplots of Fig.~\ref{fig:coherencethree}, inconsistent with our inference, we found that the orbital eccentricity of the sample clusters does not have any significant relationship with their MDs. It is also not significantly related to the ER of the sample clusters either, with only local ups and downs in the distribution between the ERs and the orbital eccentricities. This manifests that the ER may be more sensitive than the MD to the orbital eccentricities of the sample clusters. Therefore, as the sample clusters move across the Galactic disk, their crossing of different potential energy surfaces may affect their ER, but not linear (see red curve in the left panels of Fig.~\ref{fig:coherencethree}b).

The median panels of the (a) and (b) subplots of Fig.~\ref{fig:coherencethree} show the distributions between the radial forces (F$_{R}$) and the MDs of the sample clusters and their ERs, respectively. We found that the MDs of the sample clusters keep no change as the F$_{R}$ increases, with their ER fluctuating. It implies that the radial forces can not impact the MDs of the sample clusters, but may cause their ERs to exhibit fluctuations by affecting their morphology. Here, we still retain the sample effects in our consideration. Therefore, this result needs to be tested by a larger sample in the future.

Moreover, apart from the radial forces, we also explored whether vertical forces are related to the MDs and ERs of the sample clusters. However, we only detected a fluctuation of the distribution between the ERs of the sample clusters and their vertical forces (F$_{z}$), with no apparent relationship among them, as shown in the right panel of Fig.~\ref{fig:coherencethree}b. The vertical forces are independent of the MD of the sample clusters, as shown in the right panel of Fig.~\ref{fig:coherencethree}a, which suggests that the MD is not affected by the vertical forces. In addition, the fluctuation between the ER and the F$_{z}$ may be due to the insufficient sample, so we need to treat this fluctuation with caution.

\subsection{Special sample clusters}

Apart from a significant negative relationship between the ERs of the sample clusters and the numbers of their member stars, there are only some fluctuations in the distributions between the ERs and ages, spatial positions (X-axis and Y-axis), and external environmental parameters (eccentricity, F$_{R}$, and F$_{z}$). This also suggests that the external morphologies of the sample clusters must be easily impacted by the external environment, thus their ERs are sensitive to the external force environment. In this section, we aim to study the 3D spatial distributions of the sample clusters with a large ER or a small ER.

\begin{figure}
	\centering
	\includegraphics[angle=0,width=86mm]{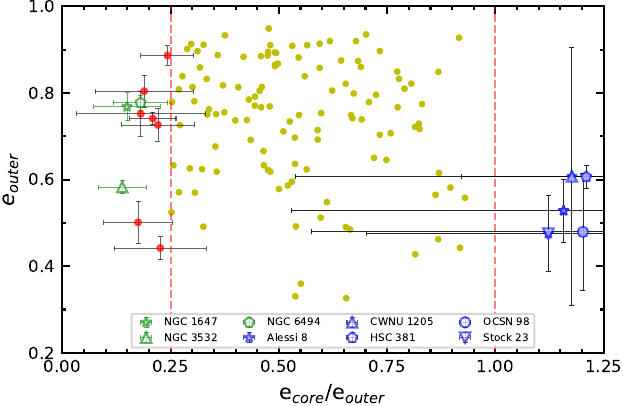}
	\caption{Distribution of the ER of sample clusters with the ellipticities of their outers. Each colored dot or shaped symbol represents one sample cluster. The black and green error bars (longitudinal and transverse) indicate the errors of the outer ellipticity and ER, respectively. The two red dashed lines are the dividing line between three regions (ER~$<$~0.25, 0.25~$\leq$~ER~$\leq$~1, and ER~$>$~1).}
	\label{fig:ratio_eouter}
\end{figure}

Figure~\ref{fig:ratio_eouter} displays the distribution between the ERs of the sample clusters and their outer ellipticities. All colored dots (yellow and red) and different kinds of symbols (marked green and blue) represent sample clusters. We can see from Fig.~\ref{fig:ratio_eouter} that most sample clusters (yellow dots) are located in 0.25~$\leq$~ER~$\leq$~1. However, several special clusters can be found in ER~$<$~0.25. Their external morphology is more elliptical than their internal morphology, which is more elliptical than most sample clusters (yellow dots). Among these sample clusters, there are seven sample clusters with ER values lower than 0.25 and its error bar extends to out of 0.25. Only three open clusters have ERs less than 0.25 with their error bars in the 0.25 range. They are marked by different kinds of symbols, which are \textit{NGC~1647} (green open pentagram), \textit{NGC~6949} (green open pentagon), and \textit{NGC~3532} (green open triangle), respectively, as shown in Fig.~\ref{fig:ratio_eouter}.

\vspace{12pt}

\begin{figure*}
	\centering
	\includegraphics[angle=0,width=60mm]{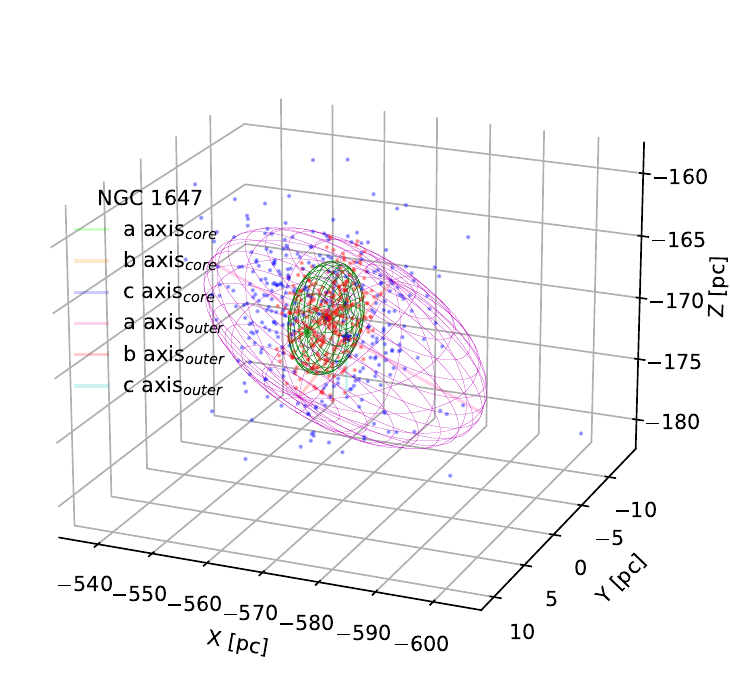}
	\includegraphics[angle=0,width=60mm]{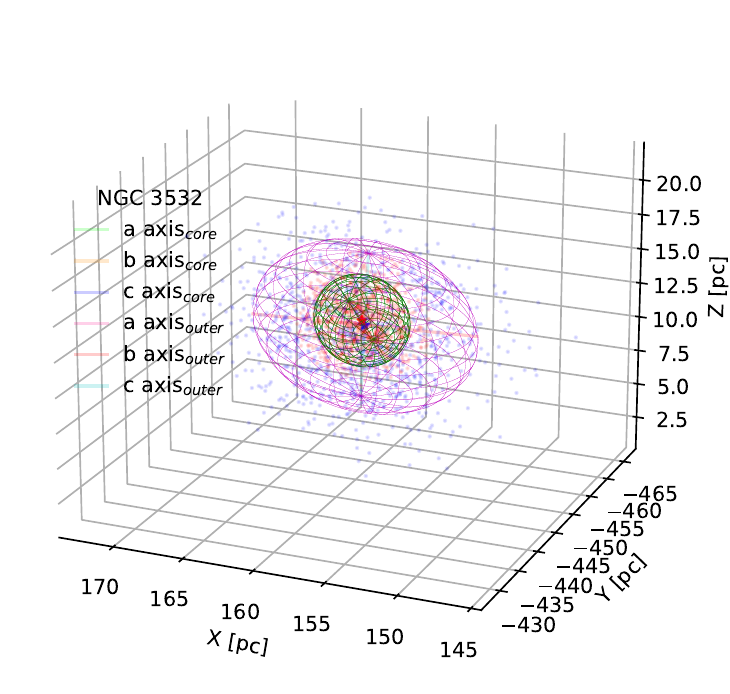}
	\includegraphics[angle=0,width=60mm]{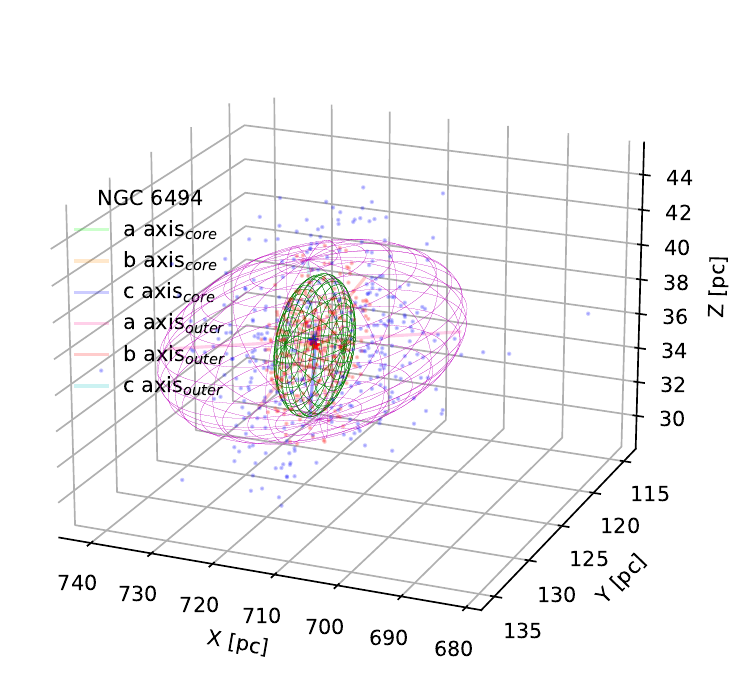}
	\includegraphics[angle=0,width=60mm]{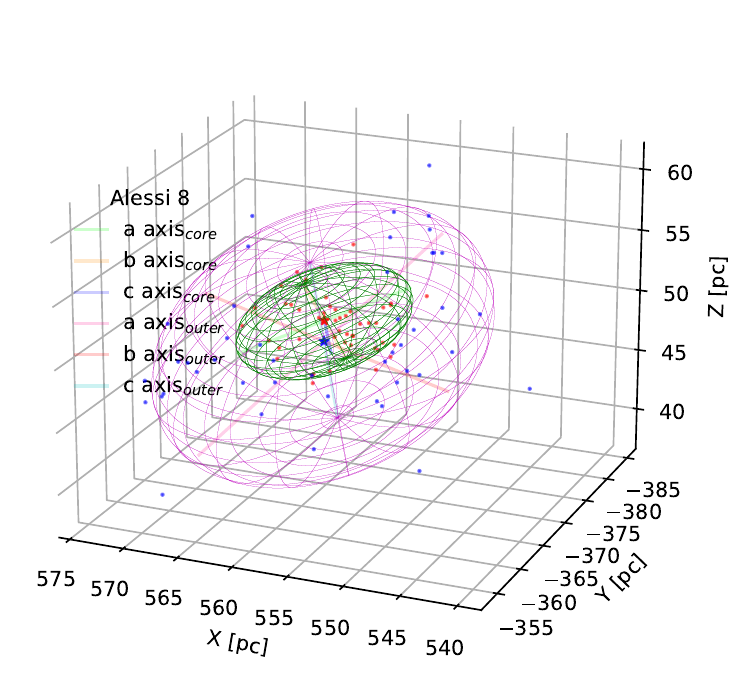}
	\includegraphics[angle=0,width=60mm]{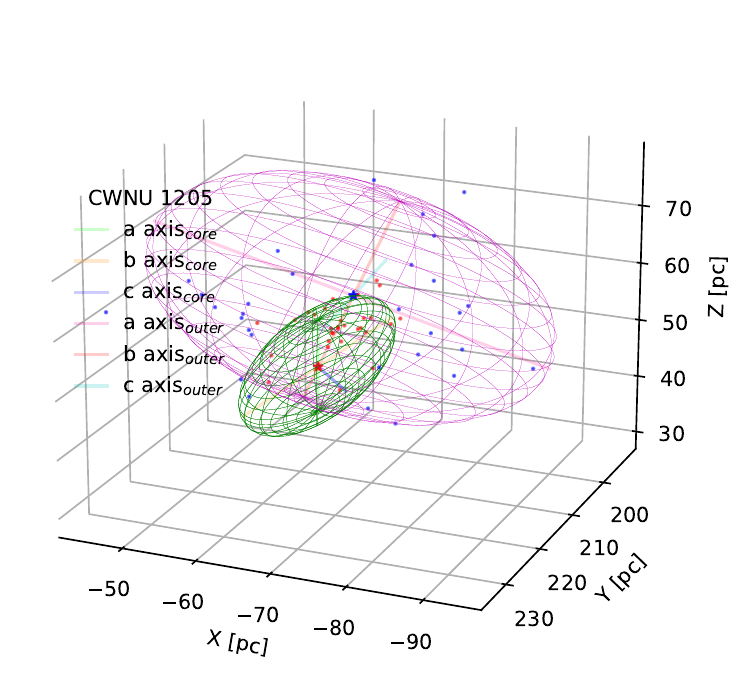}
	\includegraphics[angle=0,width=60mm]{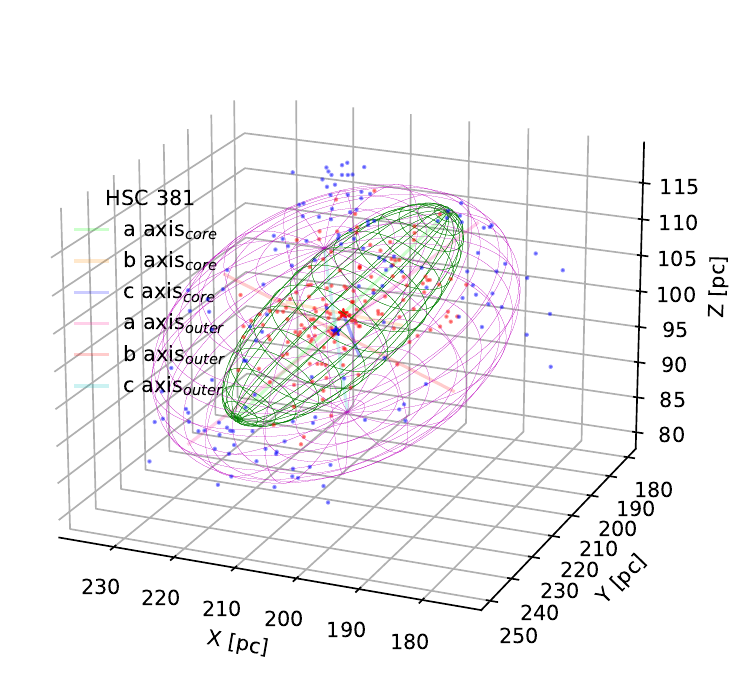}
	\includegraphics[angle=0,width=60mm]{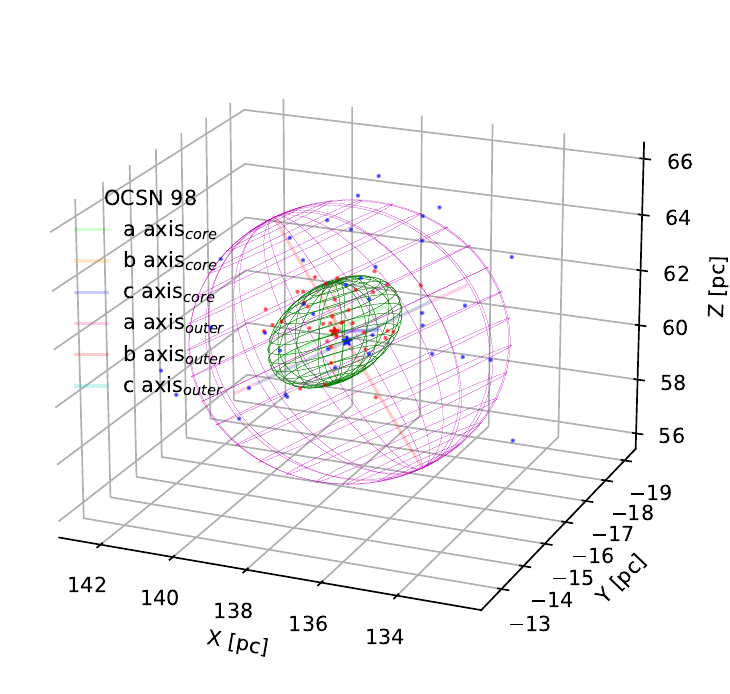}
	\includegraphics[angle=0,width=60mm]{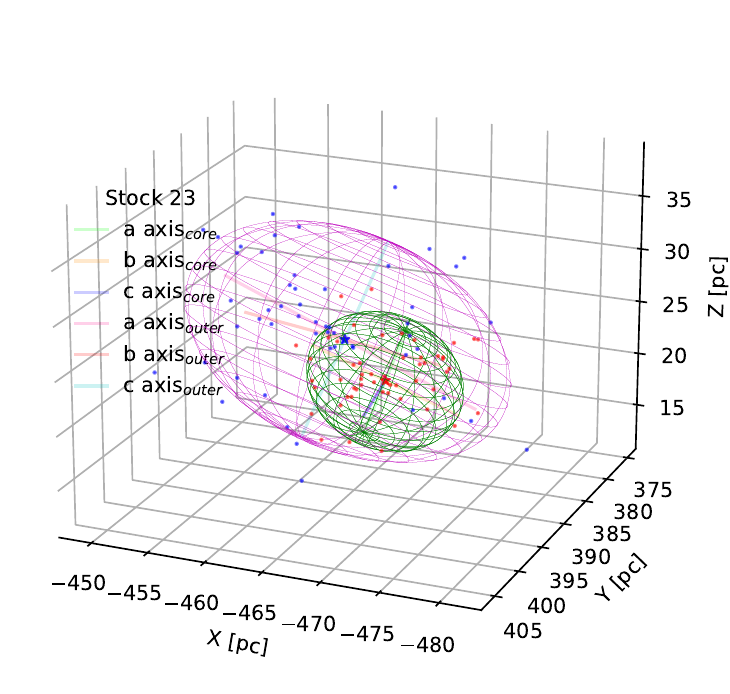}
	
	\caption{3D spatial structures of eight sample clusters (\textit{NGC~1647}, \textit{NGC~3532}, \textit{NGC~6494}, \textit{Alessi~8}, \textit{CWNU~1205}, \textit{HSC~381}, \textit{OCSN~98},  and \textit{Stock~23}) and their ellipsoidal curves. All symbols and curves are the same as those in Fig.~\ref{fig:Collinder_135_ellipsoid}}
	\label{fig:OCs_ellipsoid}
\end{figure*}

Figure~\ref{fig:OCs_ellipsoid} shows the 3D spatial distribution of eight special clusters of our sample, as well as the fitted ellipsoids of their internal and external structures. From the upper left panel of this picture, we can see the 3D spatial structure of \textit{NGC~1647}, including its internal and external structures. \textit{NGC~1647} is an intermediate-age open cluster (about~117~Myr), with 599 members with a probability more than or equal to 0.5 and 383 stars with a probability less than 0.5, and about~577~pc \citep{hunt23}. Previous research \citep[e.g.,][]{hass72, pism77, geff96, zdan05, guer11, sanc18} has focused mainly on its photometry and proper motion, whereas the study of its morphology was rarely covered. \citet[][see their Fig~7]{zdan05} reported \textit{NGC~1647} having a loose structure and elongated in the direction roughly perpendicular to our galaxy, which was based on the star counts method in the infrared $K$ passband. However, in this work, we found that it has a compact internal structure and loose external morphology, with a small ER. This means that its external region is more elliptical than its internal area. Based on the 3D spatial distribution of its external region, we speculated that it may be undergoing stellar vaporization, with escaped stars gradually becoming members of the tidal tail.

\vspace{12pt}

We can see from the upper median panel of Fig.~\ref{fig:OCs_ellipsoid}, the 3D spatial structure of \textit{NGC~3532}. It is an extremely rich cluster, located in the Carina field of the Milky Way \citep{egge81}, with an intermediate age (about~238~Myr) and about 472~pc, has 1843 members with probability more than or equal to 0.5 and 1571 stars with probability less than 0.5 \citep{hunt23}. \citet{gies81} pointed out that \textit{NGC~3532} has an extended cluster halo. In the study, it presents a spherical internal structure and a slightly elliptical external system. Meanwhile, its small MD implies a high degree of morphological stability. Due to its dense internal core, there is a high probability of two-body interaction, which allows low-mass members to transfer into the outer regions of the cluster and causes massive stars to sink into its core area. Therefore, this leads to a loose external structure of \textit{NGC~3532}.

\vspace{12pt}

\textit{NGC~6494} in our sample is also a cluster with a more elliptical external shape than internal morphology, as shown in the upper right panel of Fig.~\ref{fig:OCs_ellipsoid}. At about 713~pc and an intermediate-age (about~226~Myr) \citep{hunt23}, it displays an intermediate richness \citep{sand80}, with 465 members with membership less than 0.5 and 663 stars with membership more than or equal to 0.5 \citep{hunt23}. Resembling the 3D spatial structures of \textit{NGC~1647} and \textit{NGC~3532}, \textit{NGC~6494} has also a dense internal region and a relatively loose external structure, with a small MD.

\vspace{12pt}

Furthermore, we can also see in Fig.~\ref{fig:ratio_eouter} several sample clusters that are opposite to the above clusters. They are \textit{Alessi~8} (blue open pentagram), \textit{CWNU~1205} (blue open triangle), \textit{OCSN~98} (blue open circle), \textit{HSC~381} (blue open pentagon), and \textit{Stock~23} (blue open inverted triangle), with their ERs greater than 1, while the error bars of their ER jump out of that range. Therefore, we try to analyze the reasons for the instability of their ERs as well as the morphological stability of these clusters.

\vspace{12pt}

\textit{Alessi~8}, about~654~pc, is an intermediate age (about~101~Myr), with 93 members more than or equal to 0.5 and without members less than 0.5 \citep{hunt23}. We found from the median left panel of Fig.~\ref{fig:OCs_ellipsoid} that it has no dense internal core, with a very loose external system. This may be due to a few number of its members. In addition, given its intermediate age and present members' number, we suspected that it was probably born as a small cluster. Although it has a relatively small MD, it is likely to be dissolved by any external disruption due to its weak gravitational binding. We also found that due to its large ER, its internal morphology is more elliptical than its external structure, which indicates there is unlikely to appear a compact core in its internal region. Therefore, this structure tends to be unstable due to the difficulty of overcoming external perturbations.

\vspace{12pt}

\textit{CWNU~1205}, discovered firstly by \citet{he22}, is a young open cluster (about~20~Myr) and about 230~pc, with 60 members more than or equal to 0.5 and without members less than 0.5 \citep{hunt23}. Like \textit{Alessi~8}, \textit{CWNU~1205} has a small number of member stars. It can be seen from the central panel of Fig.~\ref{fig:OCs_ellipsoid} that its 3D spatial structure is very loose in both the inner region and the outer region. Therefore, the ellipsoid fitting errors of its inner and outer structures are relatively large, which leads to a large error in ER. For this cluster, however, its MD is larger. Combining the MD of \textit{Alessi~8} and \textit{CWNU~1205}, we speculated that the MD as an indicator of the morphological stability of sample clusters may rest on a sufficient number of member stars.

\vspace{12pt}

\textit{HSC~381}, a young ($\sim$~31~Myr) open cluster and about~313~pc, has 288 members with more than or equal to 0.5 and without members less than 0.5 \citep{hunt23}. The median right panel of Fig.~\ref{fig:OCs_ellipsoid} shows its 3D spatial structure. We can see that its internal shape is very elliptical, more than its external ellipticity, with a small MD. However, the ellipsoid fitting error of its external region is relatively large for its loose external distribution of members, and then resulting in its large error in ER. We speculated that it may gradually form a relatively clustered internal core under two-body interaction if no catastrophic event occurs. This is because it is young and has a relatively large number of member stars.

\vspace{12pt}

\textit{OCSN~98}, first discovered by \citet{qin23}, is a close ($\sim$~151~pc) and very young ($\sim$~4.5~Myr) open cluster, with 75 members more than or equal to 0.5 and without members less than 0.5 \citep{hunt23}. Interestingly, we found that it has a circular external structure and a slightly elliptical internal morphology, with a small MD, as displayed in the lower left panel of Fig.~\ref{fig:OCs_ellipsoid}. Due to its loose external system, the ellipsoid fitting error of its external region is relatively large, thus causing also its large error in ER. Furthermore, it is strange for \textit{OCSN~98} to present the circular external morphology. Because of its low height ($|$Z$|$~$<$~100~pc) and its small number of members, it gets a high probability of being stretched by the gravitational perturbation of the Galactic disk.

\vspace{12pt}

\textit{Stock~23} contains 106 members with probability more than or equal to 0.5 and without members less than 0.5, with $\sim$598~pc and young age ($\sim$~72~Myr) \citep{hunt23}. As shown in the lower right panel of Fig.~\ref{fig:OCs_ellipsoid}, we found that \textit{Stock~23} has a very loose either in its internal region or in its external area, which causes a relatively large error of its ER. It can be seen that its MD is large. Combined with a few number of its members and its large MD, it is not difficult to speculate that it is morphologically unstable.

\section{Summary}

This work is based on the biggest census catalog of open clusters and their member star catalogs from \textit{Gaia}-DR3, from which 132 open clusters with a 3D core-shell structure that fit the ellipsoidal model have been selected as our sample. Using the sample clusters' MDs and ERs to characterize their morphological coherence, we investigated the distribution of the morphological coherence (MD and ER) of the sample clusters, exploring them in relation to foundational parameters (age and member numbers), spatial positions, and an external environment. Finally, we analyzed the special samples with a small ER or a large ER. Some of the conclusions drawn from this work are as follows:

1. It suggests that although the MD error of \textit{HSC~2986} is large based on our method of dividing it into two parts according to the half-number radius, the sample cluster indeed has a large MD through GMM verification.

2. The results indicate that there is an anticorrelation of the ER of sample clusters with the number of their members, implying that the sample clusters with a vast number of members may have much more elliptical external shape than internal morphology. In addition, the result of a slight shrinking of the MD of the sample clusters with the members' number hints that the self-gravitational binding of the sample clusters plays a significant role in maintaining their morphological stability.

3. Our study demonstrates that both the MD and ER of the sample clusters are independent of their age. Nevertheless, several fluctuations of the ER on the distribution between it and the age seem to show that as the sample clusters age, their morphologies may change back and forth.

4. We found that both the MD and ER of the sample clusters are not significantly correlated with the X-axis and Y-axis. However, their ERs show some fluctuations along the X-axis, which may demonstrate that the morphologies of the sample clusters, especially their external morphologies, are susceptible to external environments at different locations.

5. The MD of the sample clusters is not significantly correlated with their orbital eccentricities nor with the radial and vertical forces on them. In addition, the ER of the sample clusters is also not significantly correlated with these covariates. However, we can still observe that the ER shows some fluctuations in its distribution with each of these covariates, which implies that the morphology of the sample clusters is sensitive to the external force environment if sample effects are not taken into account.

6. Our analysis of the 3D spatial structures of special sample clusters suggests that the number of members may be a key factor in whether the sample clusters have a dense internal region. At the same time, to regard the MD of the sample clusters as an indicator of their morphological stability should build on a certain number of member stars.

For the first time, we investigated the morphological coherence (MD and ER) of an open cluster with a 3D core-shell structure. What we can be sure of is that the morphological coherence of open clusters would certainly change in the struggle between their gravitational binding and the constant perturbations of the external environment. In future studies of cluster morphology, it will become commonplace to unravel the intertwined external environmental perturbations that affect the morphology of open clusters, as this will contribute to a clear understanding of the morphological evolution of open clusters. We acknowledg that after the submission of this work, a revised classification for the cluster catalog of \citet{hunt23} was published in \citet{hunt24}, which may impact our samples and results. In addition, we need to acknowledge that some sample clusters still have members in tidal tails, thus our fitting method is also adopted to these sample clusters with gravitationally bound and non-gravitational bound member stars. In the analysis of results, we did not distinguish between sample clusters with or without tidal tail members. In future research, we will explore these two types of open clusters separately.

\begin{acknowledgements}
	We thank the anonymous reviewer for constructive suggestions and comments, which greatly improved the quality and clarity of the paper. This work is supported by the National Natural Science Foundation of China (NSFC) under grant 12303037, the Fundamental Research Funds of China West Normal University (CWNU, No.493065), the National Key R\&D Program of China (Nos. 2021YFA1600401 and 2021YFA1600400), the National Natural Science Foundation of China (NSFC) under grant 12173028, the Chinese Space Station Telescope project: CMS-CSST-2021-A10, the Sichuan Youth Science and Technology Innovation Research Team (Grant No. 21CXTD0038), and the Innovation Team Funds of China West Normal (No. KCXTD2022-6). Qingshun Hu would like to acknowledge the financial support provided by the China Scholarship Council program (Grant No. 202308510136). Li Chen acknowledges the support from the National Natural Science Foundation of China (NSFC) through the grants 12090040 and 12090042. Jing Zhong would like to acknowledge the NSFC under grants 12073060, and the Youth Innovation Promotion Association CAS. S. Q. acknowledges the financial support provided by the China Scholarship Council program (Grant No. 202304910547). We would also like to thank Ms. Chunli Feng for touching up the language of the article. This study has made use of the \textit{Gaia} DR3, operated by the European Space Agency (ESA) space mission (\textit{Gaia}). The \textit{Gaia} archive website is \url{https://archives.esac.esa.int/gaia/}.

	Software: Astropy \citep{astr13,astr18}, Scipy \citep{mill11}, TOPCAT \citep{tayl05}, Galpy \citep{bovy15, mack18}.

\end{acknowledgements}

%%%%%%%%%%%%%%%%%%%%%%%%%%%%%%%%%%%%%%%%%%%5 REFERENCES %%%%%%%%%%%%%%%%%%%%%%%%%%%%%%%%%%%%5

%%%%%%%%%%%%%%%%% APPENDICES %%%%%%%%%%%%%%%%%%%%%	

\appendix

\section{Supplementary material}

\subsection{Ellipsoid fitting method} \label{app: ellipsoid fitting method}

The ellipsoid fitting method adopted in this study referred to the study of \citet{li04}.  The main idea is to use the least squares ellipsoid fitting to calculate the optimal fitting ellipsoid by matrix operations. The equation involved is Eq.~\ref{equation_ellisoid}, which is the general ellipsoid equation. We substitute the data of a cluster to be fitted into this equation and obtain the coefficients of the general expression for the best-fitting ellipsoid (A, B, C, D, E, F, G, H, I) via matrix operations. The coefficients can then be used to calculate the three axes (a,b,c) in Eq.~\ref{equation_Nellisoid}, which is a standard ellipsoid expression:

\begin{equation} \label{equation_ellisoid}
	Ax^{2} + By^{2}  +  Cz^{2} +  2Dxy + 2Exz + 2Fyz + 2Gx + 2Hy + 2Iz = 1
,\end{equation}

\begin{equation} \label{equation_Nellisoid}
	\frac{x^{2} }{a^{2} } + \frac{y^{2} }{b^{2} } + \frac{z^{2} }{c^{2} } = 1
.\end{equation}

With the ellipsoid fitting method, we can obtain the optimal fitting ellipsoid for any cluster to be fitted. However, the question of whether the optimal ellipsoid is a valid ellipsoid or not needs to be judged by the errors of the three axes. In this study, we calculated the errors ($\sigma_{a}$, $\sigma_{b}$, and $\sigma_{c}$) of the three axes of the fitting ellipsoid by the following Eqs.~\ref{eqsub}, \ref{eqsub1}, \ref{eqsub2}, \ref{eqsub3}, \ref{eqerror}, \ref{eqerror1}, and \ref{eqerror2}. In order to make this calculation simple, we firstly transformed Eq.~\ref{equation_Nellisoid} into the format of Eq.~\ref{eqsub3} by using three equivalent substitutions (Eqs.~\ref{eqsub}, \ref{eqsub1}, and \ref{eqsub2}). Because Eq.~\ref{eqerror} is the fitting error ($\sigma$) equation of Eq.~\ref{eqsub3}, then it can be used to derive the error ($\sigma_{k}$) of $k$ in Eq.~\ref{eqsub3}, see Eq.~\ref{eqerror1}. Finally, we can obtain Eq.~\ref{eqerror2} to calculate the error of the $a$ axis of fitting ellipsoids by combining Eqs.~\ref{eqsub2} and \ref{eqerror1}. Similarly, we also can calculate the errors of the $b$ and $c$ axes based on the above process. Here, $x{}'$, $y{}'$, and $k$ are equivalent substitutes for physical quantities (3D spatial coordinates: $x$, $y$, and $z$, and three axes: $a$, $b$, and $c$), with $i$ and $N$ denoting members of clusters fitted and the number of the clusters' members. We take the triaxial errors of the fitted ellipsoid less than the triaxial lengths, respectively, as a condition for valid ellipsoid fitting, which is also our criterion for the final determination of sample clusters:

\begin{equation} \label{eqsub}
	x{}'  = -x^{2}
,\end{equation}

\begin{equation} \label{eqsub1}
	y{}' = \frac{y^{2} }{b^{2} } + \frac{z^{2} }{c^{2} }
,\end{equation}

\begin{equation} \label{eqsub2}
	k = \frac{1}{a^{2} }
,\end{equation}

\begin{equation} \label{eqsub3}
	y{}' = kx{}' +1
,\end{equation}

\begin{equation}\label{eqerror}
	\sigma = \sqrt{\frac{\sum (y_{i}^{'}-kx_{i}^{'}-1)^{2} }{N-2} }
,\end{equation}

\begin{equation}\label{eqerror1}
	\sigma_{k} = \sigma \sqrt{\frac{N}{N\sum (x_{i}^{'} )^{2}  - (\sum x_{i}^{'}) ^{2} } }
,\end{equation}

\begin{equation}\label{eqerror2}
	\sigma_{a} = \frac{a^{3} }{2} \sigma _{k}
.\end{equation}

\subsection{Distance correction method} \label{distance correction}

This work uses the Bayesian distance correction method, which was originally systematically provided by \citet{bail15} and \citet{carr19}, then continued to be developed and used by \citet{pang21} and also adopted by other works \citep[e.g.,][]{ye21, hu23, qinm23}. The following formula mainly refers to the appendix of \citet{carr19}. Equation~\ref{posterior} is the posterior distribution function of the Bayesian distance correction model, while Eqs.~\ref{likelihood} and ~\ref{prior} are the likelihood and prior distribution functions of this model, respectively. The equations are as follows:

\begin{equation}\label{posterior}
	f( d\mid \omega) \propto f(\omega \mid d)\Pi (d)
,\end{equation}

\begin{equation}\label{likelihood}
	f(\omega \mid d) \propto e^{\frac{-(\omega -\frac{1}{d} )^{2} }{2\sigma _{\omega }^{2}  } }
,\end{equation}

\begin{equation}\label{prior}
	\Pi (d) = (1-\alpha )d^{2} e^{-\frac{d}{L} } + \frac{\alpha }{\sqrt{2\pi \sigma _{d}^{2} } }e^{-\frac{(d-d_{0})^{2} }{2\sigma _{d}^{2} } }
.\end{equation}

The introduction of important parameters in the above formulation is as follows: $\omega$ indicates each member parallax measurement, with $\sigma_{\omega }$ equal to the error bar of each member parallax measurement; $d_{0}$ denotes each cluster center, with L = 8~kpc \citep{bail15}; $\alpha$ represents the member probability; and $\sigma_{d}$ is the standard deviation of the distances of all member stars of a cluster to the cluster center, which was also uesd by \citet{pang21}, with $d$ being the corrected distance of each star. We refer to \citet{carr19, pang21} for a more detailed Bayesian distance correction method.

\begin{figure} 
	\centering
	\includegraphics[angle=0,width=62mm]{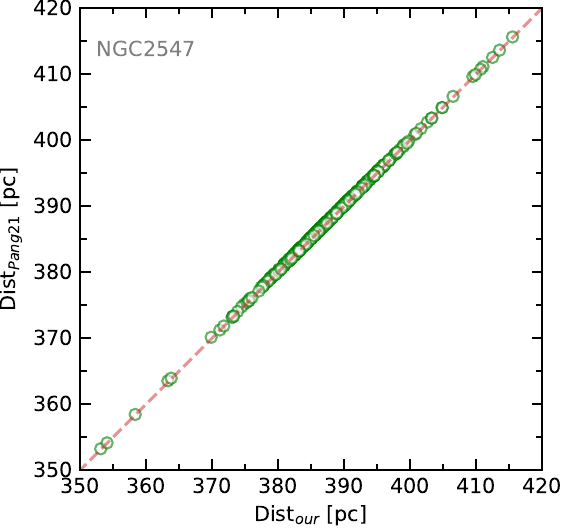}
	\caption{Validation of Bayesian distance correction by correcting distances to member stars of NGC2547. The diagram displays the distribution between the member distance corrected by \citet{pang21} and the member distance by us, both using the Bayesian distance correction method. Each open green circle in the picture represents one member star of NGC2547, with the red dashed line being a diagonal line. The member's data of NGC2547 and their corrected distances are from \citet{pang21}.}
	\label{fig:NGC2547}
\end{figure}

To verify that our Bayesian approach is correct, we tested and validated our approach with the help of previous cluster data. Since there is no published corrected distance data for cluster member stars in \citet{carr19}, we carried out this test using the corrected distance data for cluster member stars in \citet{pang21}. We downloaded and obtained the raw data of the open cluster NGC~2547 and the corrected distance of the cluster from the data published in \citet{pang21}. We then tested the raw data of the cluster with our distance correction method and compared the obtained results with the corrected distances in \citet{pang21}, as shown in Fig.~\ref{fig:NGC2547}. We can see that the results of our test are consistent with the results of \citet{pang21}. Hence, our Bayesian distance correction approach is correct and effective.

\begin{sidewaystable*}[ht]\tiny
	\caption{Overall parameters of sample clusters}\label{table:parameters}
	\centering
	\begin{tabular}{lccccccccccccccccccc}
		\hline\noalign{\smallskip}
		\hline\noalign{\smallskip}
		Name &  ra & dec & $N_{p~\geq~0.5}$  & logt &  X & Y  & Z & $d$  &  $\sigma_{d}$  &  $e_{core}/e_{outer}$ & $\sigma_{e_{core}/e_{outer}}$ & $F_{R}$ & $\sigma_{F_{R}}$ & $F_{z}$ & $\sigma_{F_{z}}$ & $E$ & $\sigma_{E}$ \\
		\hline\noalign{\smallskip}
		-	&  (degree) & (degree) & - &  - & (pc)  & (pc) & (pc) & - & - & - & - & 2$\pi$G$M_{\odot}$/$pc^{2}$ & 2$\pi$G$M_{\odot}$/$pc^{2}$ & 2$\pi$G$M_{\odot}$/$pc^{2}$ & 2$\pi$G$M_{\odot}$/$pc^{2}$ &- & - \\
		\hline\noalign{\smallskip}
		
		  ASCC 18 & 81.33 & 0.37 & 82.0 & 6.97 & -361.87 & -149.8 & -134.46 & 0.5 & 0.23 & 0.5 & 0.32 & -208.14 & 0.11 & 18.39 & 0.68 & 0.12 & 0.03 \\
		  ASCC 20 & 82.19 & 1.72 & 194.0 & 7.1 & -323.96 & -128.75 & -111.47 & 0.07 & 0.01 & 0.37 & 0.11 & -209.43 & 0.2 & 15.44 & 0.49 & 0.11 & 0.03 \\
		  ASCC 51 & 140.4 & -69.86 & 136.0 & 8.1 & 151.98 & -511.53 & -132.4 & 0.19 & 0.07 & 0.28 & 0.14 & -223.87 & 0.24 & 21.88 & 0.72 & 0.06 & 0.04 \\
		  ASCC 58 & 153.71 & -55.02 & 236.0 & 7.76 & 96.69 & -466.78 & 10.48 & 0.29 & 0.05 & 0.19 & 0.11 & -222.59 & 0.24 & -6.36 & 0.69 & 0.07 & 0.04 \\
		  ASCC 79 & 229.37 & -61.07 & 214.0 & 6.74 & 642.14 & -543.61 & -44.44 & 0.11 & 0.03 & 0.64 & 0.13 & -241.81 & 0.23 & 5.89 & 0.85 & 0.09 & 0.04 \\
		  ASCC 113 & 317.96 & 38.55 & 344.0 & 8.26 & 70.57 & 555.16 & -65.09 & 0.19 & 0.02 & 0.26 & 0.11 & -221.37 & 0.19 & 8.74 & 0.97 & 0.09 & 0.03 \\
		  ASCC 127 & 347.3 & 65.18 & 247.0 & 7.09 & -143.3 & 346.83 & 27.17 & 0.17 & 0.02 & 0.51 & 0.12 & -214.96 & 0.12 & -10.08 & 1.47 & 0.09 & 0.05 \\
		  Alessi 3 & 109.19 & -46.68 & 186.0 & 8.8 & -56.9 & -261.96 & -73.09 & 0.08 & 0.04 & 0.53 & 0.14 & -217.69 & 0.23 & 9.97 & 0.69 & 0.15 & 0.02 \\
		  Alessi 6 & 220.01 & -66.14 & 203.0 & 8.56 & 606.47 & -635.32 & -85.34 & 0.16 & 0.03 & 0.59 & 0.1 & -239.97 & 0.36 & 15.41 & 0.7 & 0.06 & 0.02 \\
		  Alessi 8 & 232.3 & -51.31 & 93.0 & 8.0 & 560.02 & -371.99 & 49.52 & 0.13 & 0.02 & 1.16 & 0.63 & -238.89 & 0.24 & -16.08 & 0.77 & 0.09 & 0.05 \\
		  Alessi 20 & 2.64 & 58.76 & 252.0 & 6.83 & -197.06 & 376.27 & -27.43 & 0.11 & 0.04 & 0.47 & 0.17 & -213.33 & 0.11 & 1.21 & 0.48 & 0.09 & 0.06 \\
		  Alessi 21 & 107.59 & -9.32 & 256.0 & 7.9 & -416.25 & -394.19 & -0.22 & 0.24 & 0.04 & 0.49 & 0.1 & -206.74 & 0.28 & -3.69 & 0.62 & 0.11 & 0.06 \\
		  Alessi 36 & 106.57 & -37.6 & 198.0 & 7.46 & -97.56 & -251.14 & -64.27 & 0.09 & 0.02 & 0.93 & 0.22 & -216.5 & 0.08 & 8.24 & 0.41 & 0.13 & 0.04 \\
		  Alessi 37 & 341.97 & 46.29 & 396.0 & 8.1 & -142.01 & 687.29 & -141.28 & 0.45 & 0.04 & 0.49 & 0.06 & -213.91 & 0.47 & 20.92 & 1.07 & 0.09 & 0.04 \\
		  BH 99 & 159.55 & -59.11 & 484.0 & 7.65 & 127.91 & -430.87 & -4.29 & 0.22 & 0.15 & 0.43 & 0.17 & -223.67 & 0.11 & -3.3 & 0.44 & 0.07 & 0.05 \\
		  Blanco 1 & 0.91 & -30.01 & 422.0 & 8.24 & 43.3 & 11.18 & -231.96 & 0.2 & 0.02 & 0.66 & 0.24 & -219.97 & 0.08 & 35.03 & 0.18 & 0.09 & 0.0 \\
		  Briceno 1 & 81.08 & 1.7 & 171.0 & 6.91 & -306.6 & -117.62 & -109.45 & 0.24 & 0.08 & 0.6 & 0.25 & -209.98 & 0.05 & 15.32 & 0.29 & 0.11 & 0.04 \\
		  CWNU 147 & 188.75 & -58.09 & 50.0 & 8.09 & 413.12 & -694.44 & 66.32 & 0.29 & 0.04 & 0.72 & 0.17 & -232.59 & 0.16 & -18.66 & 0.36 & 0.07 & 0.04 \\
		  CWNU 232 & 92.35 & -2.99 & 138.0 & 8.11 & -418.8 & -243.99 & -89.22 & 0.18 & 0.11 & 0.76 & 0.35 & -206.63 & 0.23 & 11.15 & 1.1 & 0.07 & 0.05 \\
		  CWNU 522 & 75.68 & 8.08 & 173.0 & 7.55 & -356.56 & -78.14 & -129.96 & 0.11 & 0.03 & 0.56 & 0.12 & -208.34 & 0.41 & 17.55 & 0.87 & 0.09 & 0.02 \\
		  CWNU 1032 & 300.92 & 31.85 & 231.0 & 7.97 & 138.82 & 348.83 & -0.96 & 0.22 & 0.04 & 0.77 & 0.1 & -223.95 & 0.5 & -4.08 & 1.55 & 0.08 & 0.05 \\
		  CWNU 1083 & 126.14 & -41.02 & 100.0 & 6.88 & -64.93 & -333.19 & -11.23 & 0.4 & 0.04 & 0.47 & 0.17 & -217.56 & 0.17 & -1.19 & 1.21 & 0.13 & 0.08 \\
		  CWNU 1205 & 322.2 & 70.54 & 60.0 & 7.31 & -67.99 & 215.19 & 56.41 & 0.58 & 0.43 & 1.18 & 0.64 & -217.35 & 0.26 & -14.89 & 1.15 & 0.09 & 0.05 \\
		  Collinder 69 & 83.81 & 9.89 & 753.0 & 6.72 & -376.34 & -102.35 & -82.61 & 0.14 & 0.02 & 0.53 & 0.1 & -207.99 & 0.11 & 10.56 & 0.81 & 0.11 & 0.03 \\
		  Collinder 135 & 109.41 & -36.92 & 209.0 & 7.47 & -105.05 & -274.18 & -57.55 & 0.08 & 0.02 & 0.61 & 0.21 & -216.25 & 0.11 & 7.03 & 0.47 & 0.12 & 0.04 \\
		  Collinder 140 & 110.8 & -32.01 & 203.0 & 7.24 & -162.39 & -346.12 & -52.36 & 0.18 & 0.03 & 0.43 & 0.12 & -214.46 & 0.16 & 5.9 & 0.54 & 0.13 & 0.05 \\
		  Collinder 350 & 267.06 & 1.42 & 136.0 & 8.39 & 320.74 & 161.84 & 94.41 & 0.15 & 0.04 & 0.48 & 0.2 & -230.18 & 0.19 & -23.53 & 0.54 & 0.07 & 0.02 \\
		  Collinder 463 & 27.29 & 71.8 & 454.0 & 8.17 & -518.15 & 676.89 & 140.63 & 0.12 & 0.06 & 0.32 & 0.07 & -202.66 & 0.29 & -24.22 & 0.49 & 0.13 & 0.04 \\
		  FSR 0569 & 37.82 & 72.47 & 333.0 & 6.73 & -583.96 & 687.17 & 176.36 & 0.25 & 0.03 & 0.7 & 0.11 & -200.52 & 0.42 & -27.72 & 0.5 & 0.14 & 0.04 \\
		  Gulliver 9 & 127.2 & -47.92 & 285.0 & 7.01 & -42.36 & -495.78 & -47.26 & 0.09 & 0.07 & 0.46 & 0.15 & -217.9 & 0.15 & 5.23 & 0.98 & 0.11 & 0.07 \\
		  Gulliver 20 & 273.84 & 11.44 & 126.0 & 8.41 & 317.78 & 258.61 & 95.92 & 0.07 & 0.03 & 0.67 & 0.14 & -230.17 & 0.22 & -23.77 & 0.68 & 0.06 & 0.03 \\
		  Gulliver 21 & 106.89 & -25.51 & 140.0 & 8.45 & -343.53 & -539.86 & -90.43 & 0.08 & 0.01 & 0.27 & 0.14 & -208.43 & 0.18 & 11.88 & 0.47 & 0.08 & 0.06 \\
		  HSC 381 & 271.39 & 17.77 & 288.0 & 7.49 & 206.4 & 213.97 & 101.7 & 0.09 & 0.0 & 1.21 & 0.29 & -226.26 & 0.38 & -24.03 & 1.33 & 0.07 & 0.04 \\
		  HSC 884 & 348.28 & 61.71 & 50.0 & 6.86 & -293.15 & 729.19 & 19.98 & 0.34 & 0.29 & 0.68 & 0.13 & -209.7 & 0.12 & -6.93 & 0.63 & 0.14 & 0.05 \\
		  HSC 1262 & 57.64 & 35.07 & 143.0 & 6.7 & -353.32 & 131.97 & -99.14 & 0.11 & 0.06 & 0.55 & 0.18 & -208.58 & 0.08 & 13.28 & 0.53 & 0.1 & 0.02 \\
		  HSC 1640 & 67.14 & -12.16 & 128.0 & 7.22 & -167.12 & -87.71 & -142.95 & 0.23 & 0.03 & 0.78 & 0.1 & -214.04 & 0.55 & 19.95 & 1.61 & 0.1 & 0.03 \\
		  HSC 1761 & 105.61 & -10.62 & 57.0 & 8.26 & -461.75 & -440.9 & -26.23 & 0.27 & 0.05 & 0.43 & 0.22 & -205.3 & 0.29 & 0.82 & 0.85 & 0.1 & 0.04 \\
		  HSC 1865 & 103.66 & -24.23 & 246.0 & 6.88 & -502.84 & -717.18 & -158.32 & 0.09 & 0.03 & 0.48 & 0.1 & -203.19 & 0.34 & 21.05 & 0.64 & 0.14 & 0.06 \\
		  HSC 1961 & 96.55 & -36.06 & 90.0 & 8.0 & -283.76 & -591.92 & -235.99 & 0.27 & 0.18 & 0.54 & 0.24 & -209.23 & 0.31 & 31.46 & 1.02 & 0.12 & 0.06 \\
		  HSC 2032 & 99.18 & -44.43 & 50.0 & 7.89 & -156.16 & -515.46 & -206.48 & 0.35 & 0.06 & 0.82 & 0.17 & -213.54 & 0.21 & 29.7 & 0.6 & 0.13 & 0.05 \\
		  HSC 2384 & 159.05 & -64.21 & 166.0 & 7.75 & 179.49 & -531.77 & -50.7 & 0.13 & 0.03 & 0.65 & 0.09 & -225.13 & 0.32 & 6.42 & 0.84 & 0.09 & 0.04 \\
		  HSC 2468 & 182.23 & -51.54 & 210.0 & 6.92 & 48.8 & -94.91 & 7.68 & 0.37 & 0.11 & 0.77 & 0.06 & -221.52 & 0.19 & -6.62 & 1.12 & 0.04 & 0.03 \\
		  HSC 2907 & 244.77 & -24.45 & 349.0 & 6.99 & 148.0 & -22.69 & 49.02 & 0.53 & 0.05 & 0.86 & 0.14 & -224.6 & 0.16 & -13.42 & 1.16 & 0.07 & 0.05 \\
		  HSC 2919 & 246.6 & -24.08 & 175.0 & 7.43 & 131.4 & -14.53 & 41.91 & 0.22 & 0.06 & 0.66 & 0.33 & -224.07 & 0.04 & -12.55 & 0.31 & 0.13 & 0.11 \\
		  HSC 2986 & 285.32 & -36.88 & 222.0 & 6.85 & 146.43 & -0.04 & -46.36 & 1.22 & 0.96 & 0.75 & 0.34 & -224.61 & 0.09 & 3.82 & 1.21 & 0.06 & 0.04 \\
		  Harvard 10 & 244.81 & -54.97 & 236.0 & 8.05 & 596.07 & -346.03 & -40.26 & 0.04 & 0.02 & 0.46 & 0.21 & -240.4 & 0.24 & 5.08 & 0.78 & 0.04 & 0.03 \\

		... & ...& ... & ...& ... & ... & ...& ... & ... & ... & ... & ... & ... & ... & ... & ... & ... & ...  \\
		
		\hline\noalign{\smallskip}
	\end{tabular}
	\tablefoot{Column~(1) represents the name of sample clusters. Columns~(2) and (3) denote the right ascension and declination of the sample clusters, respectively, with these data being taken from \citet{hunt23}. Column~(4), and (5) refers to the number of the sample clusters' members with probability more than or equal to 0.5 and their age, respectively, also from \cite{hunt23}. Column~(6), (7), and (8) indicate the X, Y, and Z of the sample clusters; Columns~(9) and (10) are the MD and its error with Columns~(11) and (12) being the ER and its error; Columns~(13) and (14) are the radial force and its error with Columns~(15) and (16) being the vertical force and its error; Columns~(17) and (18) are the orbital eccentricity and its error. This is a machine-readable table, and the complete table can be found in the CDS.}
	\flushleft
\end{sidewaystable*}

\end{document}